\documentclass{article}
\usepackage[letterpaper, margin=1in]{geometry}
\usepackage{authblk}

\usepackage{comment}
\usepackage{tikz}
\usetikzlibrary{matrix,arrows,decorations.pathmorphing,patterns,positioning}
\usepackage{amssymb}
\usepackage{ifthen}
\usepackage{color}
\usepackage{colortbl}
\usepackage{rotating}
\usepackage{multirow}
\usepackage{multicol}
\usepackage{url}

\usepackage{pgfplots}

\usepackage{wrapfig}
\usepackage{subfigure}
\usepackage{array}
\usepackage{framed}
\usepackage{multirow}

\usepackage{epsfig}
\usepackage{amsmath}
\usepackage{amsfonts}
\usepackage{amssymb}
\usepackage{algorithmicx}

\usepackage[normalem]{ulem}

\usepackage{ifthen}
\usepackage{boxedminipage}
\usepackage{fancyvrb}

\usepackage{soul}
\usepackage{graphicx}
\usepackage{graphics}
\usepackage{moreverb}
\usepackage{rotating}
\usepackage{enumerate}

\usetikzlibrary{arrows}

\usepackage{listings}
\usepackage{xcolor}

\usepackage{pdflscape}

\newcommand{\gemm}{{\sc gemm} }

\newcolumntype{I}{!{\vrule width 1.5pt}}
\newlength\savedwidth

\newboolean{IsWide}
\setboolean{IsWide}{true}

\newcommand{\FlaPartition}[2]{
\ifthenelse{\boolean{IsWide}}{{\bf partition } \hspace{-1em} #1 \hspace{-1em} #2}
{{\bf partition } \+ \\ #1 \+ \\ #2 \- \-}
}

\newcommand{\FlaRepartition}[2]{
\ifthenelse{\boolean{IsWide}}{{\bf repartition } \hspace{-1em} #1 \hspace{-1em} #2}
{{\bf repartition } \+ \\ #1 \+ \\ #2 \- \-}
}

\newcommand{\FlaContinue}[1]{
\ifthenelse{\boolean{IsWide}}{{\bf continue with } #1
}
{{\bf continue with } \+ \\ #1 \-
}
}

\newcommand{\blocksize}{1}

\newcommand{\repartitionings}{
\begin{minipage}[t]{2in}
\ \\
\ \\
\ \\
\end{minipage}
}

\newcommand{\repartitionsizes}{ \hspace{ 1.25in} }

\newcommand{\WSrepartition}{
\begin{minipage}[t]{2in}
\ifthenelse{ \equal{\blocksize}{1} }{}
{%
\ifthenelse{ \equal{\blocksize}{2} }{~}
{\bf Determine block size $ \blocksize $} \\
}
{\bf Repartition}
\begin{tabbing}
in \= in \= \+ \kill
\repartitionings \+ \\
{\bf where } \hspace*{-2ex} \repartitionsizes 
\end{tabbing}
\end{minipage}
}

\newcommand{\WSrepartitionNarrow}{
\begin{minipage}[t]{2.15in}
\ifthenelse{ \equal{\blocksize}{1} }{}
{%
\ifthenelse{ \equal{\blocksize}{blank} }{~}
{\bf Determine block size $ \blocksize $} \\
}
{\bf Repartition}
\begin{tabbing}
i \= i \= \+ \kill
\repartitionings \+ \\
{\bf where } \hspace*{-2ex} \repartitionsizes 
\end{tabbing}
\end{minipage}
}

\newcounter{mycounter}

\newcommand{\Gemm}{\mbox{\sc Gemm}}

\begin{document}


\title{Automating the Last-Mile for High Performance Dense Linear Algebra}
\author[1]{Richard M. Veras (rveras@cmu.edu)}
\author[1]{Tze Meng Low (lowt@cmu.edu)}
\author[2]{Tyler M. Smith (tms@cs.utexas.edu)}
\author[2]{Robert van de Geijn (rvdg@cs.utexas.edu)}
\author[1]{Franz Franchetti (franzf@cmu.edu)}
\affil[1]{Department of Electrical and Computer Engineering, Carnegie Mellon University}
\affil[2]{Department of Computer Science, The University of Texas at Austin}
\footnotetext{
This work was sponsored by the DARPA PERFECT program under agreement
HR0011-13-2-0007. The content, views and conclusions presented in this
document do not necessarily reflect the position or the policy of DARPA or the
U.S. Government. No official endorsement should be inferred.

Author's addresses: R. Veras, T. M. Low {and}  F. Franchetti, Department of
Electrical and Computer Engineering, Carnegie Mellon University, 5000 Forbes
Avenue, Pittsburgh, Pennsylvania 15213; R. van de Geijn, T. Smith, The
University of Texas at Austin, Austin Texas, 78705.
}

\maketitle

\begin{abstract}
High performance dense linear algebra (DLA) libraries often rely on a general
matrix multiply (Gemm) kernel that is implemented using assembly or with
vector intrinsics. The real-valued Gemm kernels provide the overwhelming
fraction of performance for the complex-valued Gemm kernels, along with the
entire level-3 BLAS and many of the real and complex LAPACK
routines. Achieving high performance for the Gemm kernel translates into a
high performance linear algebra stack above this kernel. However, it is a
monumental task for a domain expert to manually implement the kernel for every
library-supported architecture. This leads to the belief that the craft of a
Gemm kernel is more dark art than science. It is this premise that drives the
popularity of autotuning with code generation in the domain of DLA.

This paper, instead, focuses on an analytical approach to code generation of
the Gemm kernel for different architecture, in order to shed light on the
details or "voo-doo" required for implementing a high performance Gemm
kernel. We distill the implementation of the kernel into an even smaller
kernel, an outer-product, and analytically determine how  available SIMD
instructions can be used to compute the outer-product efficiently. We codify
this approach into a system to automatically generate a high performance SIMD
implementation of the Gemm kernel. Experimental results demonstrate that our
approach yields generated kernels with performance that is competitive with
kernels implemented manually or using empirical search.

\end{abstract}

\section{Introduction}
High performance dense linear algebra libraries (e.g. the Level 3
Basic Linear Algebra Subroutines (BLAS)~\cite{BLAS3}, and
LAPACK~\cite{LAPACK}) are often written using a high performance
general matrix-matrix multiplication routine
(\gemm)~\cite{Goto:2008:HIL3,poorman_journal}.  This {\gemm} routine
is typically custom-written by a domain expert for a particular
architecture.  These domain expert are often affectionately called
``assembly-ninjas'', for two main reasons. Firstly, the routines
written by these experts are frequently written using low-level
languages, i.e. assembly, that are architecture-specific. More
importantly, the process by which these experts implement these
architecture-specific routines are often unknown, or considered too
intricate and architecture-specific to be worth modeling.  In some
sense, the work performed by these experts are more akin to some form
of dark art than science.

It is for that reason that many of the productivity improvement tools
such as auto-tuning and code-generation~\cite{PHiPAC97,ATLAS,AuGem} are
focused on generating code that are auto-tuned around a small
black-box kernel that is still implemented by the expert.  The main
idea is for the auto-tuner/code-generator to find the loop ordering,
and blocking sizes surrounding the optimized kernel that ``best''-suit
the targeted architecture, while being agnostic to the details of the
hand-coded implementations.  The smaller architecture-specific kernel
reduces the implementation effort of the expert, thus allowing more
high performance kernels, each targetting different architectures, to
be implemented.

\begin{figure}
\begin{center}
\includegraphics[scale=0.5]{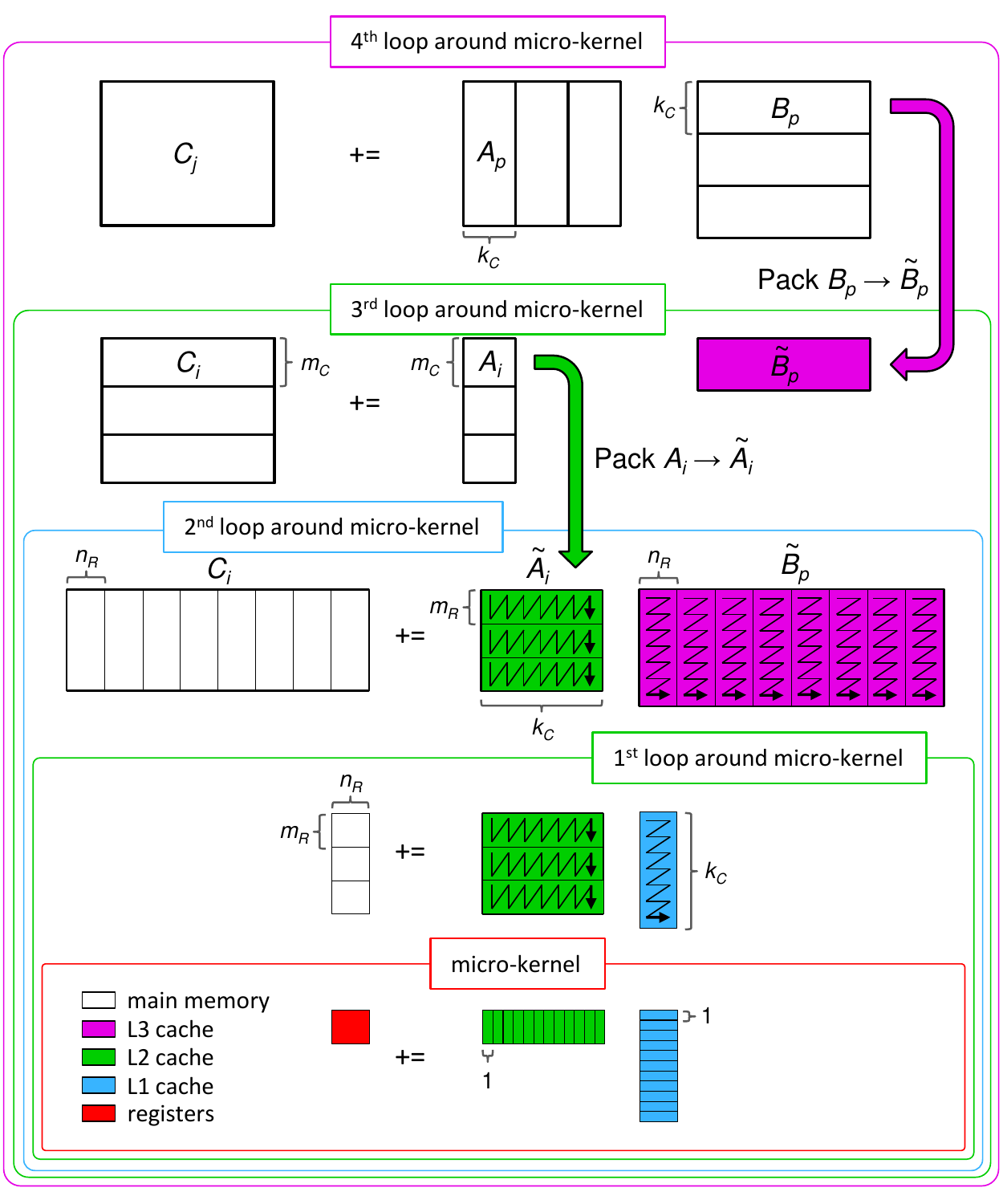}
\end{center}
\caption{The GotoBLAS approach for matrix multiplication as refactored in
BLIS. Blocked arrows represent explicit data packing, and thin arrows
represent the data layout in after packing.}
\label{fig:BLIS}
\end{figure}

Around the turn of the century, Kazushige Goto revolutionalized the
way matrix-matrix multiplication is implemented with the GotoBLAS
approach~\cite{Goto:2008:AHP}.  Specifically, Goto demonstrated that
data movement and computation for computing {\gemm} can be {\emph
systematically} orchestrated in a specific manner, as depicted in
Figure~\ref{fig:BLIS}, that does not change as we shift between
architectures. In Goto's implementations, a small {\em macro-kernel}
needs to be custom-implemented.  The BLIS framework~\cite{BLIS1}
refactored Goto's algorithm exposing additional loops so that only a
much smaller micro-kernel needs to be customized for a new
architecture, with all the other parts implemented in the C
programming language.

A great deal of work has focused on the generation of cache resident small
matrix kernels such as LGen~\cite{LGEN,LGEN2,LGEN3} and the Built-to-Order
BLAS \cite{BTOBLAS}. For many scientific and engineering application there is
a need for linear algebra operations on cache resident data which these
kernel code generators solve. However, because these generators require that
the inputs reside in the lowest level of cache they are not appropriate for
larger out-of-cache {\gemm} operations.

Prior to this paper, the point at which science became art is in the
implementation of the micro-kernel itself. With this paper, we push
science one step further, exposing the system behind the
implementation of the micro-kernel. Specifically, we make the
observation that a micro-kernel is inherently implemented as a loop
around outer-products. The execution of that loop can be thought of as
a first stage consisting of one or more iterations that suffer a
certain degradation in performance as hardware resources are being
consumed, a second stage during which each iteration involves
computation with data that already exists in registers that can mask
data movement to be used in this iteration or future iterations, and a
third stage during which final computations are performed and results
are flushed from the system. The performance of the micro-kernel is
primarily determined by how efficient the second stage is and that is
where we therefore focus our efforts. We codify this effort as a model based
code generation system for micro-kernels, thus automating the last-mile for
producing high performance {\gemm} routines.

\section{Anatomy of a {\Gemm} micro-kernel}

BLIS is a framework for rapidly instantiating the BLAS using the GotoBLAS
(now maintained as OpenBLAS\cite{OpenBLAS}) approach~\cite{Goto:2008:AHP}, and
it is one of the most efficient expert-tuned implementations of the BLAS.  The
GotoBLAS approach essentially performs loop tiling
\cite{Padua:1986:ACO:7902.7904} and packing of data for different layers~\cite{GMH:92}
within the cache hierarchy in a specific manner to expose an inner
kernel. The specific loop tiling strategy in GotoBLAS has been shown
to work well on many modern CPU architectures. BLIS extends these
ideas, and tiles the loops in the GotoBLAS inner kernel to expose an
even smaller {\Gemm} kernel, the {\em micro-kernel}, and showed that
high performance is attained when this micro-kernel is
optimized~\cite{BLIS2}.


\subsection{The micro-kernel}
The micro-kernel is a small matrix multiplication that implements $ C
+\hspace{-5pt}= A B$, where $C$ is a $m_r \times n_r$ matrix, while
$A$ and $B$ are micro-panels of size $m_r \times k_c$ and $k_c \times
n_r$ respectively. In addition, because of the packing of $A$ and $B$
prior to the invocation of the micro-kernel, it can be assumed that
$A$ is stored in a contiguous block of memory in column-major order
while $B$ is contiguously stored in row-major order.  Since the
micro-kernel is a small {\gemm} kernel, the micro-kernel can be
described, using compiler terminology, as a {\gemm} kernel computed
using a triply-nested loop of the KIJ or KJI variant.

Within the BLIS framework, it can also be assumed that $m_r, n_r \gg
k_c$. In addition, we assume that the bounds of the loops (i.e. $k_c$,
$m_r$, and $n_r$) are determined analytically using the models
from~\cite{BLIS4}.

{\bf Computing the micro-kernel.}
Mathematically,  the micro-kernel is computed by first partitioning $A$ into
columns and  $B$ into rows. The output $C$ is then computed in the following
manner:
\begin{eqnarray*}
C & +\hspace{-5pt}= & \large ( \begin{array}{c|c|c} a_0 & \hdots & a_{k_c-1}\end{array} \large ) \left ( \begin{array}{c}b_0^T \\ \hline \vdots \\ \hline b_{k_c-1}^T\end{array}\right) \\
   & +\hspace{-5pt}= & \sum^{k_c-1}_{i = 0} a_i b_i^T,
\end{eqnarray*}
where the fundamental computation is now
\[
C +\hspace{-5pt}= a_ib_i^T,
\]
a single {\em outer-product}, and our task is to compute the
outer-product multiple times, each time with a new column and row from
$A$ and $B$, in as efficient a manner as possible.

\subsection{Decomposing the outer-product}
Focusing on a single outer-product, $C += ab^T$, we can decompose the
outer-product further by performing loop tiling of $C$.  This, in
turn, will require us to block the columns of $A$ and rows of $B$ into
sub-columns and sub-rows of conformal lengths respectively, as
follows:
\[
C \rightarrow
\left(\begin{array}{c|c|c}
C^{0,0} & C^{0,1} & \hdots \\ \hline
C^{1,0} & C^{1,1} & \hdots \\ \hline
\vdots & \vdots & \ddots \\
\end{array}\right)
a \rightarrow
\left(
\begin{array}{c}
a^{0} \\ \hline
a^{1} \\ \hline
\vdots \\
\end{array}
\right)
b \rightarrow
\left(
\begin{array}{c}
b^{0} \\ \hline
b^{1} \\ \hline
\vdots \\
\end{array}
\right)
\]

\begin{figure}
\begin{center}
\includegraphics[scale=0.4]{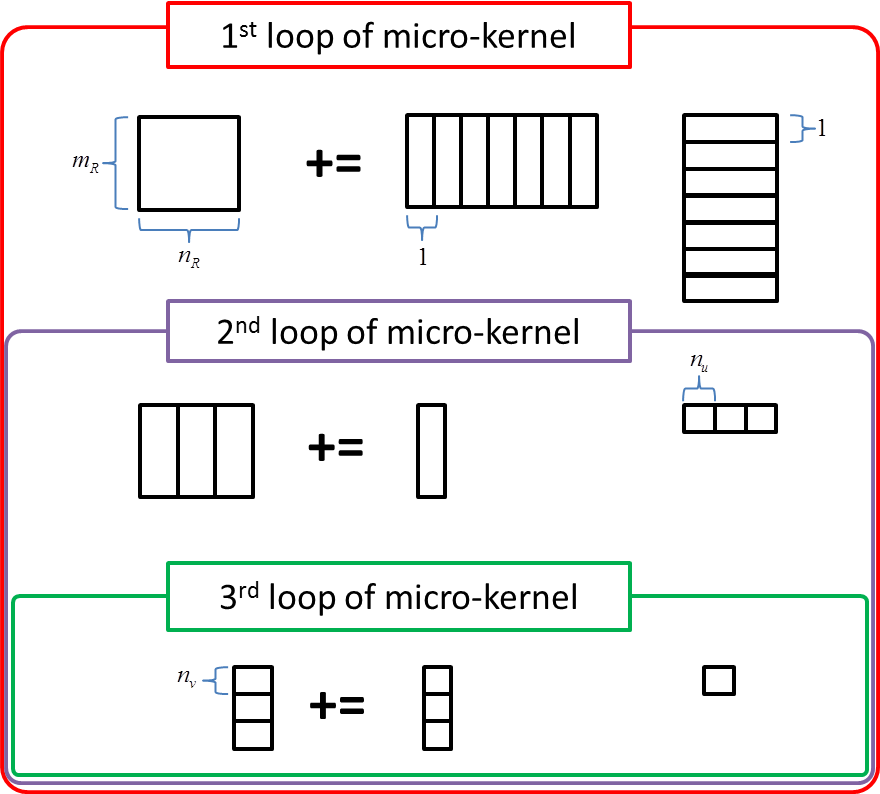}
\end{center}
\caption{ An additional three loops are introduced after decomposing
  the BLIS micro-kernel into smaller outer-product kernels of size
  $m_v \times n_u$. These set of loops would replace the micro-kernel
  shown in Figure~\ref{fig:BLIS}.}
\label{fig:rank1}
\end{figure}

In this case, the outer-product is decomposed into a smaller unit of
computation, which we will term as {\em Unit Update}, which computes
\[
C^{i,j} += (a^{i})^T b^{j},
\]
where $C^{i,j}$ is now a $m_v \times n_u$ matrix, and the subvectors
of $a^{i}$ and $b^{j}$ are vectors of length $m_v$ and
$n_u$. Decomposing the outer-product into smaller unit updates now allow
us to determine how the outer-product can be computed with the
available instructions on the targeted architectures. In the case
where $m_v = n_u = 1$, each unit update computes a single element in the
matrix. This means that the outer-product, $C$, is computed
element-wise, can be performed using the following 
code:
\[
\begin{array}{l}
\mbox{for }i = 0; : m_r/m_v - 1 \\
\quad  \mbox{for }j = 0; : n_r/n_u - 1 \\
\quad \quad    C^{i,j} +\hspace{-5pt}= (a^{i})^Tb^{j}, 
\end{array}
\]
where $b^{j}$ is streamed from the L1 cache, and $a^{i}$
is loaded into the registers from the L2 cache.
Alternatively, interchanging the two loops yields,
\[
\begin{array}{l}
\mbox{for }j = 0; : n_r/n_u - 1 \\
\quad \mbox{for }i = 0; : m_r/m_v - 1 \\
\quad \quad    C^{i,j} +\hspace{-5pt}= (a^{i})^Tb^{j},
\end{array}
\]
where each iteration of the inner-most loop requires new values of 
$a^{i}$ to be loaded from the L2 cache. In either case, we can replace
the micro-kernel block in Figure~\ref{fig:BLIS} with the new diagram in 
Figure~\ref{fig:rank1}.

The interesting case is when $m_v~\neq~1$ and/or $n_u~\neq~1$. In this
case, each unit update is a smaller outer-product. By selecting
appropriate values of $m_v$ and $n_u$, we gain the flexibility of
mapping the computation of the unit update ($C^{i,j}$) to the
availability of vector / single-instruction-multiple data (SIMD)
instructions available on modern architecture. This flexibility of
mapping also yields us a family of algorithms that compute the
outer-product, which is the kernel within the micro-kernel we are
trying to optimized.

\section{Identifying Outer-Product Kernels}
Recall that the micro-kernel is essentially a loop around multiple
outer-products. In addition, each outer-product can be
further decomposed into smaller unit updates of size $m_v \times
n_u$.  In this section, we discuss how the values of $m_v$ and $n_u$
are determined by the instructions available on the desired
architecture.

\subsection{The building blocks: SIMD instructions}
The key to high performance is to use Single Instruction Multiple Data (SIMD)
vector instructions available on many of the modern processors. We assume
that all computation involves double precision arithmetic and that each
vector register can store $v$ double precision floating point numbers, where
$v$ is a power of two. In addition, we assume that the following classes of
vector instructions are available:
%
\begin{enumerate}
\item Vector {\bf Stores.} Vector store instructions write
  all $v$ elements of a vector register to memory.
\item Vector {\bf Load.} Load
  instructions read $u$ unique elements of data from memory, where
  $u~\le~v$ and $u$ is a power of two.  An element is considered
  unique if it resides in a unique memory address.  In cases where
  $u~<~v$, each of the $u$ unique elements are duplicated $v/u$ times.

  We assume that all elements loaded by a single Load instruction are
  within $v$ memory addresses of each other\footnote{This implies that
  our framework do not handle vector gather instructions. However,
  these Gather/Scatter instructions are not required in dense linear
  algebra kernels, as matrices are often packed for locality.}.
  Prefetches are considered Load instructions.

\item Vector {\bf Shuffles.} Shuffle Instructions reorders and/or duplicates
  the elements in a vector register. We restrict ourselves to only
  instructions that be represented by a $n_v \times n_v$ matrix where each row
  contains exactly $n_v-1$ {\em zeros} and   a single {\em one}, but each
  column may contain multiple {\em ones}. In addition, we assume that the
  number of {\em ones} in each column is a power of two.

\item Vector {\bf Computation.}  It is assumed instructions that
  perform computation on vector registers are element-wise
  operations. This means that, given vector registers, $\texttt{reg\_a}$
  and $\texttt{reg\_b}$, the output is of the form:
\[
\texttt{reg\_a} \: \mbox{op} \: \texttt{reg\_b} =
\left(
\begin{array}{rcl}
\alpha_0&op_0&\beta_0 \\
\alpha_1&op_1&\beta_1 \\
&\vdots \\
\alpha_{v-1}&op_{v-1}&\beta_{v-1}
\end{array}
\right),
\]
where $op_i$ and $op_j, i \neq j$ may be different binary
operators. The result of the computation may be stored in one of the
input registers, or a third vector register.

\item {\bf Composite Instructions.}
  Some instructions -- we will call them Composite Instructions -- can be
  viewed as a combination of some of the previous   three types of
  instruction.  For example, the instruction,
  \begin{center}
  {\texttt{vfmadd231pd, reg,     reg, mem{1to8}} }
  \end{center}
   instruction on the  Xeon Phi can be expressed as a Load   instruction,
  followed by a broadcast (Shuffle) instruction, followed   by a fused
  multiply-add (Computation) instruction.
\end{enumerate}

\subsection{Mapping unit updates to SIMD instructions}

Given the available SIMD instructions, one possible size of a single
unit update is for $m_v = v$ and $n_u = 1$, where $v$ is the size of a
SIMD register.  This means that $v$ values from $a$, and a single
value of $b$ is loaded into two SIMD registers, \texttt{reg\_a}
and \texttt{reg\_b}. In addition, we know that the loaded value of $b$
is duplicated $v$ times because $n_u~<~v$. Computing
with \texttt{reg\_a} and \texttt{reg\_b} will yield a single unit
update of size $v \times 1$. A single outer-product of $m_r \times n_r$
can then be computed through $m_r/m_v \times n_r/n_u$  multiple unit
updates as shown in Figure~\ref{fig:broadcast_algo}.

\begin{figure}
{\small
\begin{tabular}{ccc}
\texttt{reg\_a} & \texttt{reg\_b} & registers storing $C^{i,j}$ \\
\begin{minipage}{0.5in}
\[
\begin{array}{|c|} \hline
\alpha_0 \\
\alpha_1 \\
\alpha_2 \\
\alpha_3 \\ \hline
\end{array}
\]
\end{minipage}
&
\begin{minipage}{0.5in}
\[
\begin{array}{|c|} \hline
\beta_0 \\
\beta_0 \\
\beta_0 \\
\beta_0 \\ \hline
\end{array}
\]
\end{minipage}
&
\begin{minipage}{1in}
\[
\begin{array}{|>{\columncolor{olive!20}}c|} \hline
\chi_{00} \\
\chi_{10} \\
\chi_{20} \\
\chi_{30} \\ \hline
\end{array}
\begin{array}{|c|} \hline
\chi_{01} \\
\chi_{11} \\
\chi_{21} \\
\chi_{31} \\ \hline
\end{array}
\begin{array}{|c|} \hline
\chi_{02} \\
\chi_{12} \\
\chi_{22} \\
\chi_{32} \\ \hline
\end{array}
\begin{array}{|c|} \hline
\chi_{03} \\
\chi_{13} \\
\chi_{23} \\
\chi_{33} \\ \hline
\end{array}
\]
\end{minipage} \\
\begin{minipage}{0.5in}
\[
\begin{array}{|c|} \hline
\alpha_0 \\
\alpha_1 \\
\alpha_2 \\
\alpha_3 \\ \hline
\end{array}
\]
\end{minipage}
&
\begin{minipage}{0.5in}
\[
\begin{array}{|c|} \hline
\beta_1 \\
\beta_1 \\
\beta_1 \\
\beta_1 \\ \hline
\end{array}
\]
\end{minipage}
&
\begin{minipage}{1in}
\[
\begin{array}{|c|} \hline
\chi_{00} \\
\chi_{10} \\
\chi_{20} \\
\chi_{30} \\ \hline
\end{array}
\begin{array}{|>{\columncolor{olive!20}}c|} \hline
\chi_{01} \\
\chi_{11} \\
\chi_{21} \\
\chi_{31} \\ \hline
\end{array}
\begin{array}{|c|} \hline
\chi_{02} \\
\chi_{12} \\
\chi_{22} \\
\chi_{32} \\ \hline
\end{array}
\begin{array}{|c|} \hline
\chi_{03} \\
\chi_{13} \\
\chi_{23} \\
\chi_{33} \\ \hline
\end{array}
\]
\end{minipage} \\
\begin{minipage}{0.5in}
\[
\begin{array}{|c|} \hline
\alpha_0 \\
\alpha_1 \\
\alpha_2 \\
\alpha_3 \\ \hline
\end{array}
\]
\end{minipage}
&
\begin{minipage}{0.5in}
\[
\begin{array}{|c|} \hline
\beta_2 \\
\beta_2 \\
\beta_2 \\
\beta_2 \\ \hline
\end{array}
\]
\end{minipage}
&
\begin{minipage}{1in}
\[
\begin{array}{|c|} \hline
\chi_{00} \\
\chi_{10} \\
\chi_{20} \\
\chi_{30} \\ \hline
\end{array}
\begin{array}{|c|} \hline
\chi_{01} \\
\chi_{11} \\
\chi_{21} \\
\chi_{31} \\ \hline
\end{array}
\begin{array}{|>{\columncolor{olive!20}}c|} \hline
\chi_{02} \\
\chi_{12} \\
\chi_{22} \\
\chi_{32} \\ \hline
\end{array}
\begin{array}{|c|} \hline
\chi_{03} \\
\chi_{13} \\
\chi_{23} \\
\chi_{33} \\ \hline
\end{array}
\]
\end{minipage} \\
\begin{minipage}{0.5in}
\[
\begin{array}{|c|} \hline
\alpha_0 \\
\alpha_1 \\
\alpha_2 \\
\alpha_3 \\ \hline
\end{array}
\]
\end{minipage}
&
\begin{minipage}{0.5in}
\[
\begin{array}{|c|} \hline
\beta_3 \\
\beta_3 \\
\beta_3 \\
\beta_3 \\ \hline
\end{array}
\]
\end{minipage}
&
\begin{minipage}{1in}
\[
\begin{array}{|c|} \hline
\chi_{00} \\
\chi_{10} \\
\chi_{20} \\
\chi_{30} \\ \hline
\end{array}
\begin{array}{|c|} \hline
\chi_{01} \\
\chi_{11} \\
\chi_{21} \\
\chi_{31} \\ \hline
\end{array}
\begin{array}{|c|} \hline
\chi_{02} \\
\chi_{12} \\
\chi_{22} \\
\chi_{32} \\ \hline
\end{array}
\begin{array}{|>{\columncolor{olive!20}}c|} \hline
\chi_{03} \\
\chi_{13} \\
\chi_{23} \\
\chi_{33} \\ \hline
\end{array}
\]
\end{minipage} \\
\end{tabular}
}
\caption{SIMD computation of a $4\times4$ outer-product using four unit updates
of size $4 \times 1$. Shaded register represents the register that is being updated during the particular stage of computation.}
\label{fig:broadcast_algo}
\end{figure}

Alternatively, a different algorithm emerges when $m_v = v$ and $n_u =
2$.  The difference is that \texttt{reg\_b} would contain two unique
values, each duplicated $v/2$ times. After the first computation is
performed, the values in \texttt{reg\_b} has to be shuffled before the
next computation can be performed.  This computation-shuffle cycle has
to be repeated at least $n_u-1$ times in order to compute a single
unit update. Pictorially, this is shown in Figure~\ref{fig:shuffle2}.
The astute reader will recognized that we could have chosen to
shuffle the values in \texttt{reg\_a} without significantly changing
the computation of the unit update.

\begin{figure}
{\small
\begin{tabular}{ccc}
\multicolumn{3}{l}{Computing first $4\times2$ unit update}\\
\texttt{reg\_a} & \texttt{reg\_b} & registers storing $C$ \\
\begin{minipage}{0.5in}
\[
\begin{array}{|c|} \hline
\alpha_0 \\
\alpha_1 \\
\alpha_2 \\
\alpha_3 \\ \hline
\end{array}
\]
\end{minipage}
&
\begin{minipage}{0.5in}
\[
\begin{array}{|c|} \hline
\beta_0 \\
\beta_1 \\
\beta_0 \\
\beta_1 \\ \hline
\end{array}
\]
\end{minipage}
&
\begin{minipage}{1in}
\[
\begin{array}{|>{\columncolor{olive!20}}c|} \hline
\chi_{00} \\
\chi_{11} \\
\chi_{20} \\
\chi_{31} \\ \hline
\end{array}
\begin{array}{|c|} \hline
\chi_{01} \\
\chi_{10} \\
\chi_{21} \\
\chi_{30} \\ \hline
\end{array}
\begin{array}{|c|} \hline
\chi_{02} \\
\chi_{13} \\
\chi_{22} \\
\chi_{33} \\ \hline
\end{array}
\begin{array}{|c|} \hline
\chi_{03} \\
\chi_{12} \\
\chi_{23} \\
\chi_{32} \\ \hline
\end{array}
\]
\end{minipage} \\
\\
\multicolumn{3}{l}{After shuffles}  \\
\begin{minipage}{0.5in}
\[
\begin{array}{|c|} \hline
\alpha_0 \\
\alpha_1 \\
\alpha_2 \\
\alpha_3 \\ \hline
\end{array}
\]
\end{minipage}
&
\begin{minipage}{0.5in}
\[
\begin{array}{|c|} \hline
\beta_1 \\
\beta_0 \\
\beta_1 \\
\beta_0 \\ \hline
\end{array}
\]
\end{minipage}
&
\begin{minipage}{1in}
\[
\begin{array}{|c|} \hline
\chi_{00} \\
\chi_{11} \\
\chi_{20} \\
\chi_{31} \\ \hline
\end{array}
\begin{array}{|>{\columncolor{olive!20}}c|} \hline
\chi_{01} \\
\chi_{10} \\
\chi_{21} \\
\chi_{30} \\ \hline
\end{array}
\begin{array}{|c|} \hline
\chi_{02} \\
\chi_{13} \\
\chi_{22} \\
\chi_{33} \\ \hline
\end{array}
\begin{array}{|c|} \hline
\chi_{03} \\
\chi_{12} \\
\chi_{23} \\
\chi_{32} \\ \hline
\end{array}
\]
\end{minipage} \\
\\
\multicolumn{3}{l}{Computing second unit update}\\
\begin{minipage}{0.5in}
\[
\begin{array}{|c|} \hline
\alpha_0 \\
\alpha_1 \\
\alpha_2 \\
\alpha_3 \\ \hline
\end{array}
\]
\end{minipage}
&
\begin{minipage}{0.5in}
\[
\begin{array}{|c|} \hline
\beta_2 \\
\beta_3 \\
\beta_2 \\
\beta_3 \\ \hline
\end{array}
\]
\end{minipage}
&
\begin{minipage}{1in}
\[
\begin{array}{|c|} \hline
\chi_{00} \\
\chi_{11} \\
\chi_{20} \\
\chi_{31} \\ \hline
\end{array}
\begin{array}{|c|} \hline
\chi_{01} \\
\chi_{10} \\
\chi_{21} \\
\chi_{30} \\ \hline
\end{array}
\begin{array}{|>{\columncolor{olive!20}}c|} \hline
\chi_{02} \\
\chi_{13} \\
\chi_{22} \\
\chi_{33} \\ \hline
\end{array}
\begin{array}{|c|} \hline
\chi_{03} \\
\chi_{12} \\
\chi_{23} \\
\chi_{32} \\ \hline
\end{array}
\]
\end{minipage} \\
\\
\multicolumn{3}{l}{After shuffles} \\
\begin{minipage}{0.5in}
\[
\begin{array}{|c|} \hline
\alpha_0 \\
\alpha_1 \\
\alpha_2 \\
\alpha_3 \\ \hline
\end{array}
\]
\end{minipage}
&
\begin{minipage}{0.5in}
\[
\begin{array}{|c|} \hline
\beta_3 \\
\beta_2 \\
\beta_3 \\
\beta_2 \\ \hline
\end{array}
\]
\end{minipage}
&
\begin{minipage}{1in}
\[
\begin{array}{|c|} \hline
\chi_{00} \\
\chi_{11} \\
\chi_{20} \\
\chi_{31} \\ \hline
\end{array}
\begin{array}{|c|} \hline
\chi_{01} \\
\chi_{10} \\
\chi_{21} \\
\chi_{30} \\ \hline
\end{array}
\begin{array}{|c|} \hline
\chi_{02} \\
\chi_{13} \\
\chi_{22} \\
\chi_{33} \\ \hline
\end{array}
\begin{array}{|>{\columncolor{olive!20}}c|} \hline
\chi_{03} \\
\chi_{12} \\
\chi_{23} \\
\chi_{32} \\ \hline
\end{array}
\]
\end{minipage}
\end{tabular}

}
\caption{SIMD computation of a $4\times4$ outer-product using two unit updates
of size $4 \times 2$. As there are multiple (2) unique values, vector
shuffles must be performed to compute each unit update. Shaded
registers denotes output register for the current stage of
computation.}
\label{fig:shuffle2}
\end{figure}

\subsection{A family of outer-product algorithms}
What we have learnt from the two algorithms described previously are
 the following:
\begin{itemize}
\item[--] The size of a single unit update can be determined by the number of 
unique values loaded into registers \texttt{reg\_a}
and \texttt{reg\_b}.
\item[--] When there are more than one unique values in registers \texttt{reg\_a} and \texttt{reg\_b}, the number of computation-shuffles required is the 
minimum of $m_v$ and $n_u$.
\item[--] Loading more unique values into \texttt{reg\_b} reduces the 
number of Loads of $b$ from the L1 cache, at the cost of increasing
the number of shuffles required to compute the unit update.
\end{itemize}

Given that we chose not to shuffle \texttt{reg\_a}, this means that
there are $\log_2(v)+1$ different ways of picking $n_u$, i.e. the
number of unique elements loaded into \texttt{reg\_b} (while still
being a power of 2)\footnote{ In practice, there are less than
$\log_2(v)+1$ ways in which data can be loaded into the registers, as
the Load instructions for a particular $n_u$ value on the targeted
architecture may not be available. For example, the $4\times2$ unit update
cannot be implemented on most x86 architectures. The limited
shuffle instructions also serve to limit the 
combinatorial explosion in implementations.}.  For a given choice of
$n_u$, the different ways in which the data in \texttt{reg\_b} should
be shuffled yields different implementations for the unit update. By
accumulating the instructions for computing all $m_r/m_v \times
n_r/n_u$ unit updates, different sets of instructions, or {\em
instruction mix}, describing different implementations of the
outer-product can be obtained.

Recall that the different number of loaded unique elements results in
different number of loads and shuffle stages required. On different
architectures, the cost (in term of latency) of loads and shuffles may
differ, which suggests a need for a means to estimate the cost of 
computing with a set of instruction mix.

\section{Selecting Outer-product Kernels}
Having derived a family of algorithms to compute the outer-product, we
need to select one of these algorithms to implement. We build a model
of the architecture and then rely on queuing theory to select and
implement the kernel with the highest throughput. 

\subsection{A model architecture}
On most modern architectures, there are a number of pipelines where
each pipeline processes a subset of the entire instruction set
architecture (ISA). In addition, there are a fixed number of
functional units on each architecture, and each functional unit is
connected to each pipelines. When the instructions are sent
into the system, they are processed by the different execution
pipelines. The code is computed when the instructions have been
retired, or exited the pipelines.

\begin{figure}
\begin{center}
\includegraphics[scale=0.3]{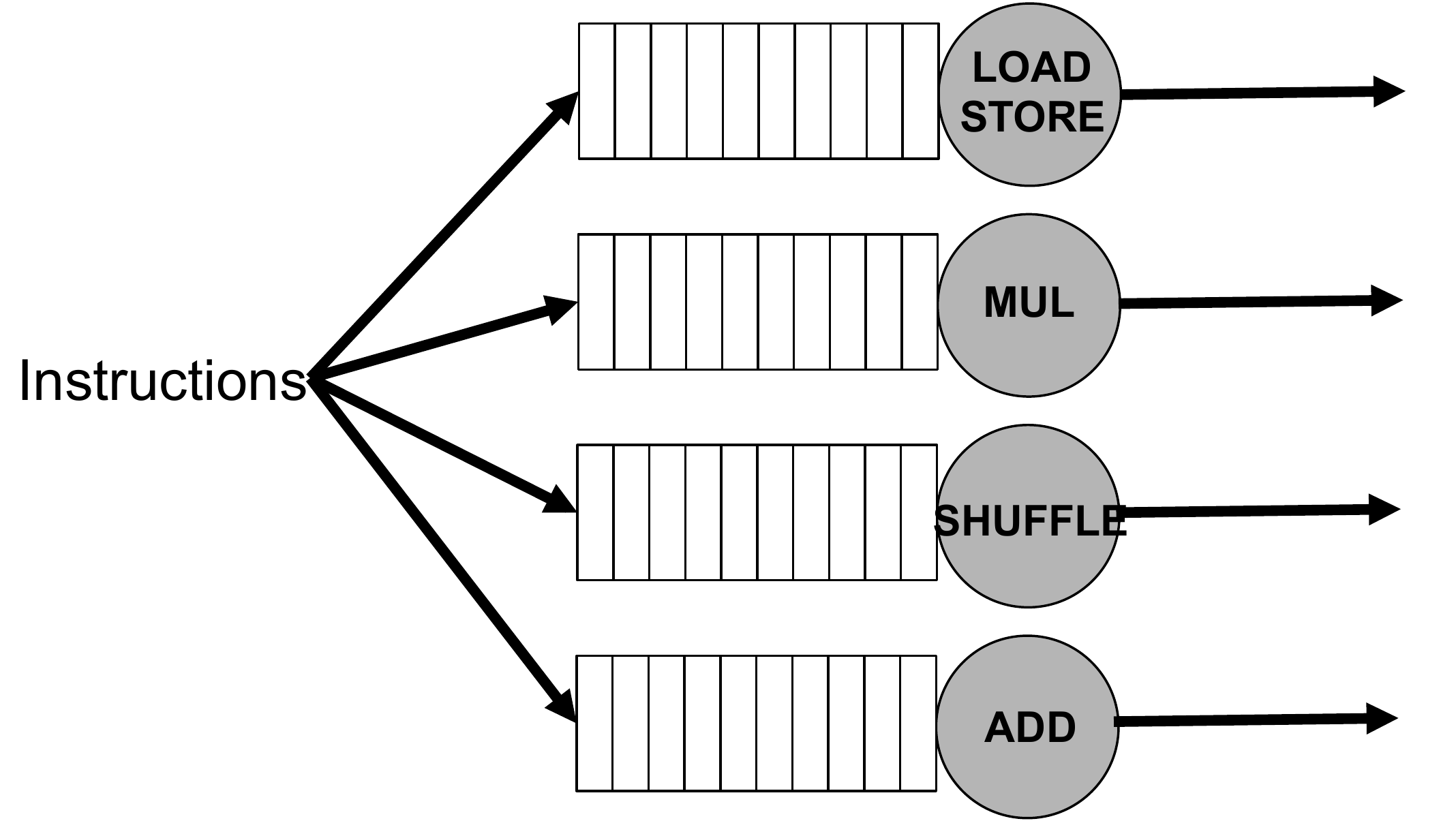}
\end{center}
\caption{Model of a subset of the Intel Sandy Bridge architecture, showing
only the floating point addition and multiplication units, the load/store
units and the vector shuffle units. Instructions enter the pipelines, and
when all instructions required for computing the outer-product leave their
respectively pipeline, the outer-product is computed.}
\label{fig:queues}
\end{figure}

Such anarchitecture can be modelled as a series of parallel
queues and servers, where pipelines and functional units are modelled
as queues and servers respectively. Instructions are jobs that are
queued in the appropriate pipelines until processed. For
instructions that require multiple functional units, we treat them as
multiple independent jobs.  A model of the Sandy Bridge architecture
relevant to the computation of the outer-product kernel is shown in
Figure~\ref{fig:queues}.

 By viewing the architecture as queues and servers, we can leverage
 queuing theory to analytically compute the throughput of computing a
 single iteration of the micro-kernel at steady state. 

 \subsection{Little's Law}
 Little's Law~\cite{little_law} states that in a steady state the
 expected number of jobs ($L$) waiting for a server in a system is
 given by:
 \[
 L = \lambda W,
 \]
 where $\lambda$ and $W$ are the average arrival rate of new jobs, 
 and average time spent in the system (waiting and processing time)
 of a job, respectively. Rearranging the above equation, 
 we obtain
 \begin{equation}
 \label{eqn:little}
 \lambda = \frac{L}{W},
 \end{equation}
 which gives us the throughput of a particular queue with an average of $L$
 jobs, each taking an average of $W$ unit of time. 

  The overall throughput of the system for computing the outer-product
  can then be estimated using
\begin{eqnarray*}
 \lambda_{\mbox{outer product}} &=& \min(\frac{1}{T_i})\quad \forall_i \mbox{pipeline},\\
&=&\min(\frac{\lambda_{i}}{L_i})
\end{eqnarray*}
where $\lambda_i$ is the throughput of pipeline $i$,  $L_i$ is the
number of instructions from the instruction mix that has been assigned
 to the pipeline. Essentially, the throughput for an outer-product is the 
inverse of the time it takes for the instruction mix to clear the pipeline
of the lowest throughput.

 \subsection{Estimating throughput}

Consider the instruction mix required to compute the $ 4~\times~4 $
outer-product shown in Figure~\ref{fig:broadcast_algo} being executed
on the model Sandy Bridge architecture described in
Figure~\ref{fig:queues}. The instruction mix contains a single
Load of a vector of $a$, four Loads (with duplication) of elements
from $b$, four multiplies and four adds (Computation)
instructions. All instructions will be sent to their respective
pipelines. In addition, another four job items are also sent to the
pipeline connected to the shuffle functional unit. This is because the
Load of element from $b$ on the SandyBridge is a Composite
instruction, that comprises of two instructions, a Load and a Shuffle.

Based on documentations from the hardware manufacturers~\cite{inteloptimize}, we
know that the throughputs of the Shuffle and Computation instructions
are $1$ per cycle, and Loads have a throughput of $2$ per cycle. This
means that the estimated throughput of the system is
\begin{eqnarray*}
 \lambda_{\mbox{outer product}} &=& 
\min(\frac{\lambda_{\mbox{load}}}{L_{\mbox{load}}},
\frac{\lambda_{\mbox{add}}}{L_{\mbox{add}}},
\frac{\lambda_{\mbox{mul}}}{L_{\mbox{mul}}},
\frac{\lambda_{\mbox{shuffle}}}{L_{\mbox{shuffle}}} ) \\
&= &\min(\frac{2}{5}, \frac{1}{4}, \frac{1}{4}, \frac{1}{4}) \\
& =& 0.25 \mbox{ outer-product per cycle}
\end{eqnarray*}
However, these throughput values are based on the assumption that the
instructions in all queues are fully pipelined, and independent.  When
the instructions in the pipelines are not independent, this implies
that the latency of executing that instruction cannot be hidden. As
such the throughput of the pipeline drops. This can happen when one
instruction has to be computed before dependent instructions can be
processed.  What this means is that the pipeline has to stall, thus
increasing the average waiting time of that pipeline.

The new average waiting time for the stalled pipeline can be esimated as 
\[
W = L + {n*k},
\]
where $n$ is the number of dependent instructions and $k$ is the
latency of the instruction. Using this new value of $W$, 
we can then compute the effective
throughput using Little's Law (Equation~\ref{eqn:little}). 

\subsection{Dealing with dependencies across pipelines}
 The authors of~\cite{BLIS4} proposed an analytical model for sizing
 the micro-kernels, where the values of $m_r$ and $n_r$ are sized such
 that all computations within a single iteration, i.e. the computation
 required to compute a single outer-product are independent. Hence, by
 adopting the BLIS analytical model presented in~\cite{BLIS4}, we know
 that all the computation instructions are independent, and can be
 pipelined without introducing bubbles into the pipelines. 

  To overcome other dependencies between other instructions, we can
  perform loop unrolling~\cite{Padua:1986:ACO:7902.7904} and software
  pipelining~\cite{software_pipeline} during code generation to
  identify and schedule independent instructions.

\section{Generating the {\Gemm} micro-kernel}

 \begin{figure} 
   \includegraphics[clip, trim=0cm 5cm 0cm 5cm,scale=.4]{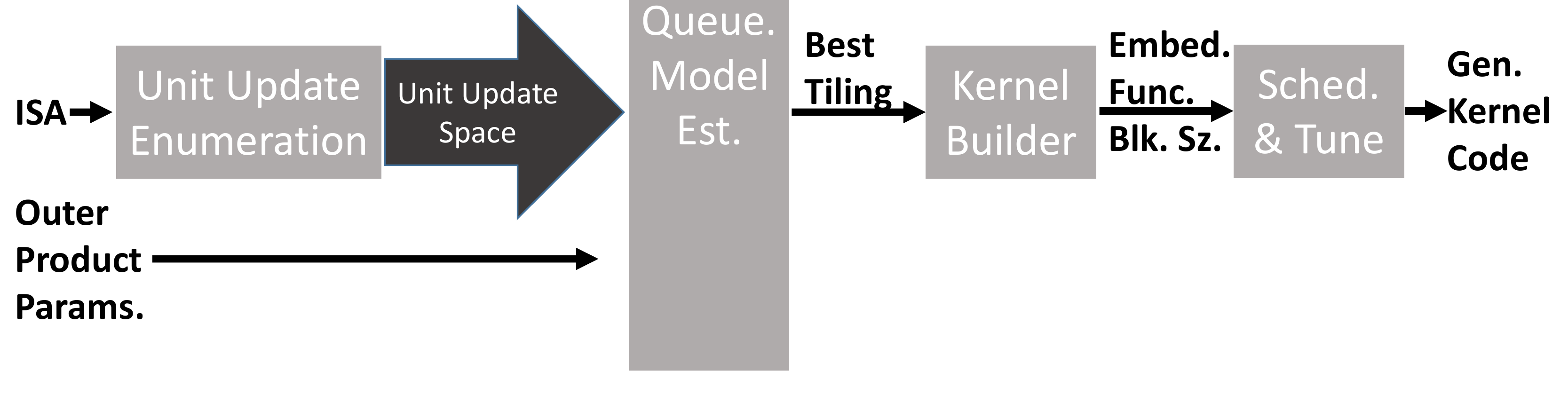}
     \caption{Our outer product kernel generation work flow produces expert
       level performing code by: enumerating a space of implementations,
       modeling their expected performance, selecting the top candidate and
       translating that into efficient code.}
     \label{fig:workflow}
 \end{figure}

In the previous section, we described a mathematical technique for
determining the most efficient mix of instructions for implementing the
outer-product kernel on a given architecture. We use queuing theory to
estimate the sustained throughput of a potential implementation without
empirical measurements. The end goal is transforming this highly parallel mix
of instructions and implementing an efficient kernel which sustains the
predicted throughput.

In this section we describe our code generator, which takes as its input an
outer-product instruction mix, and outputs a high performance kernel. By
using the more parallel outer-product formulation over an inner(dot)-product
formulation, the resulting mix contains a large number of independent
instructions. Using these independent instructions, or instruction level
parallelism (ILP), the latency of these instruction are easily hidden through
the use of static scheduling, enabling the kernel to achieve the performance
predicted by the model. Thus, the main functions of our python-based code
generator is to capture the desired instruction-mix in a template, statically
schedule the instruction mix, and emit ANSI C code in a manner that preserves
the static instruction schedule. In these three steps our generator produces a
high performance kernel that approaches the theoretical performance of an
instruction mix as determined by our queueuing theory model.

\subsection{A Work Flow for Kernel Generation}

\begin{figure}
\includegraphics[clip, trim=1cm 5cm 2cm 5cm,scale=.35]{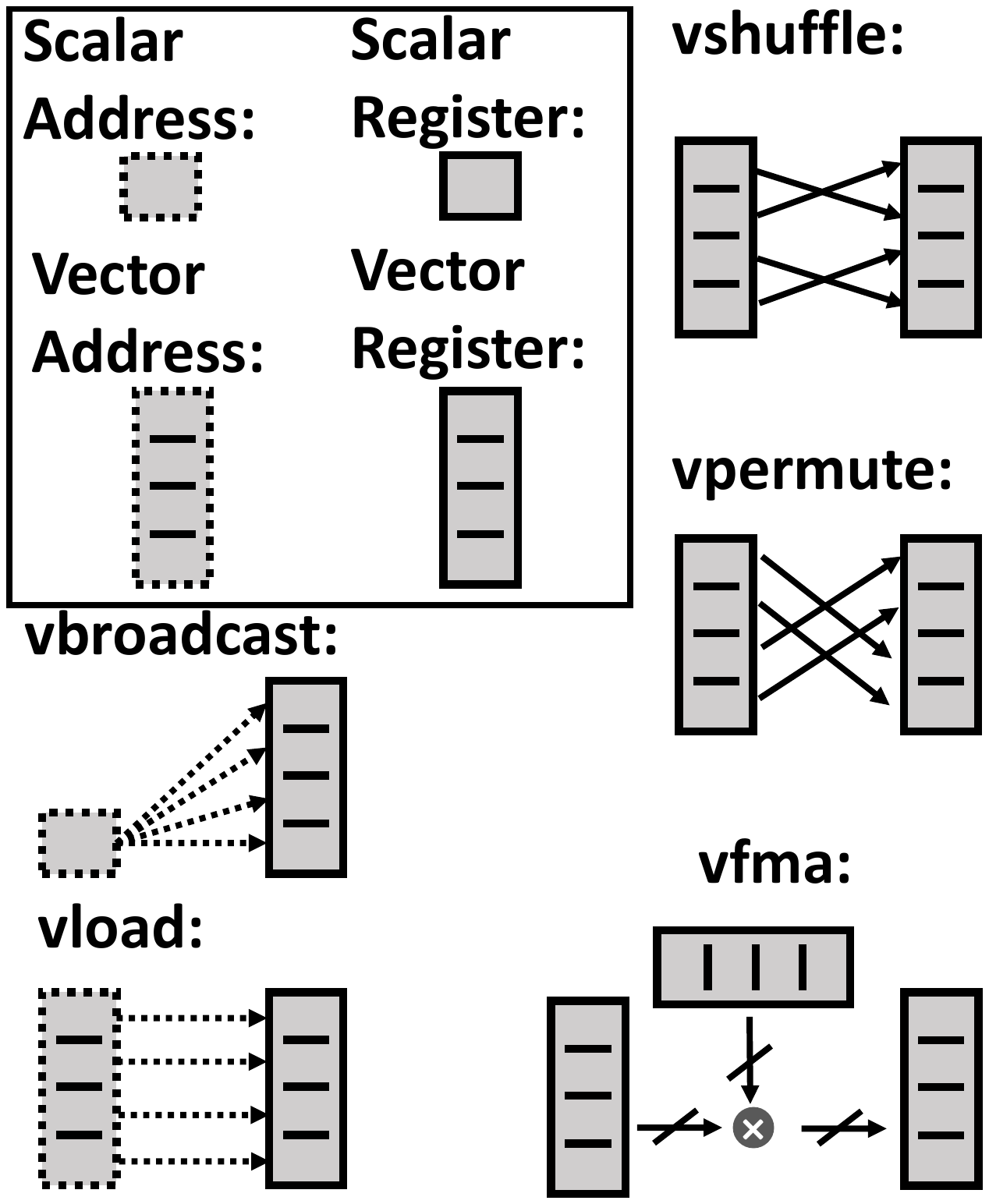}
     \caption{These cartoons illustrate the SIMD Vector instructions that are
       considered for outer-product kernel generation.}
     \label{fig:scalar_vector_registers}
     \label{fig:simd_instruction_examples}
 \end{figure}

\begin{figure} 
   \center
\includegraphics[clip, trim=5cm 2cm 4cm 3cm,scale=.4]{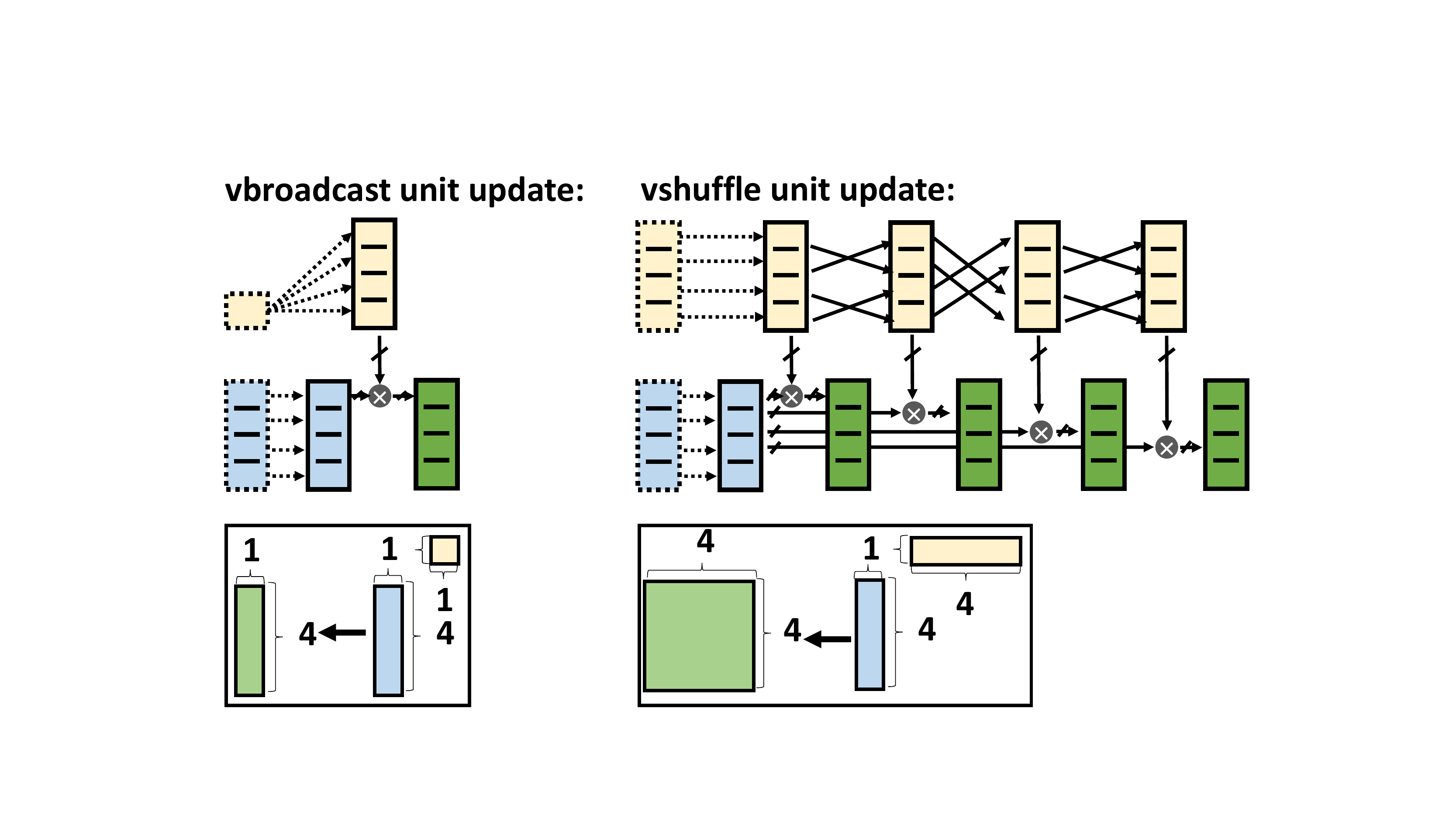}
   \caption{In this figure we show the smallest unit updates (or small
     outer products) that can be constructed from our base set of SIMD
     instructions. }
 \label{fig:components}
 \end{figure}

\begin{figure} 
   \center
\includegraphics[scale=.4]{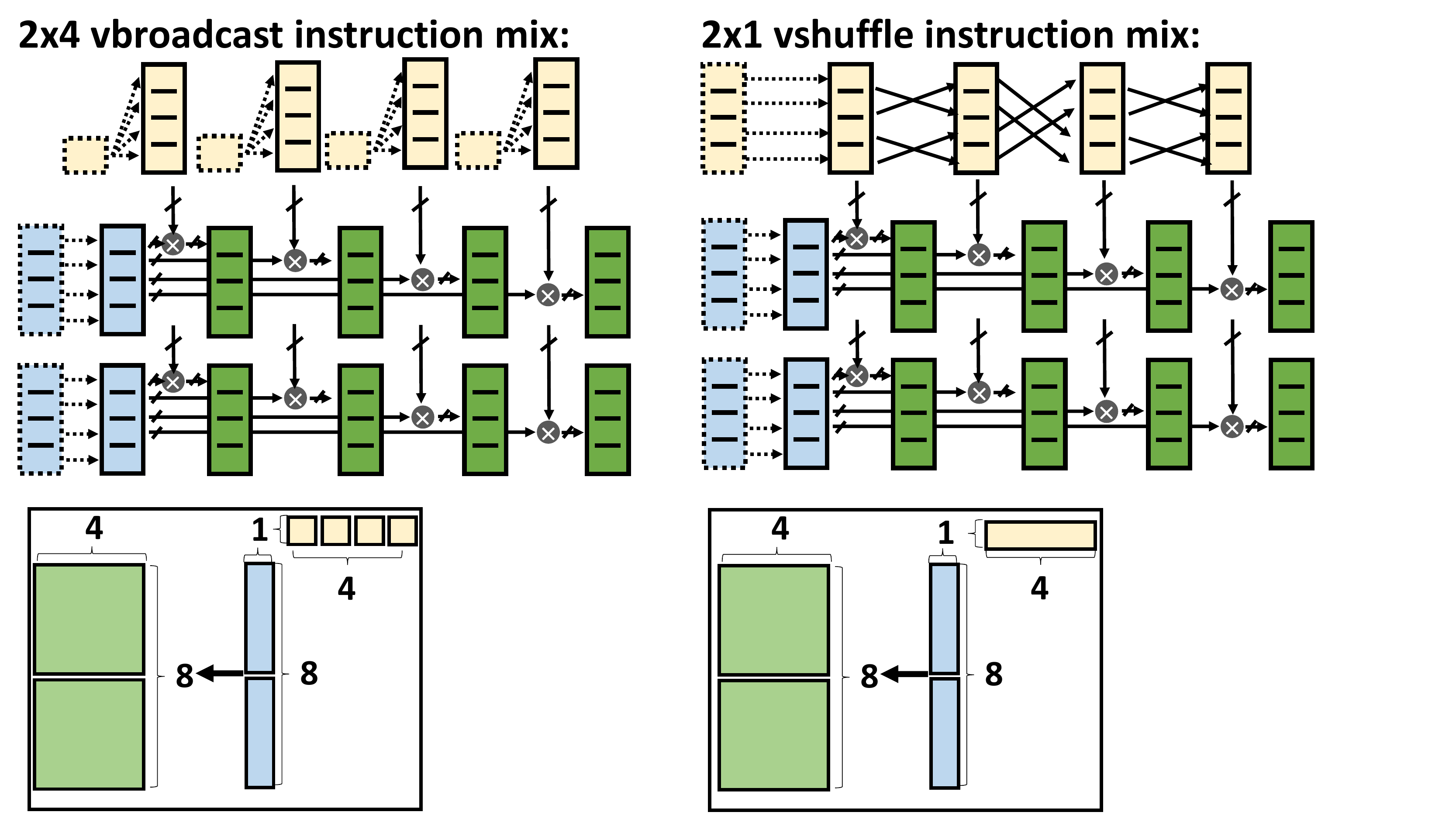}
\caption{Given the unit updates in Figure~\ref{fig:components} we can
  enumerate two possible implementations of a $m_r \times n_r = 8 \times 4$
  outer-product. On the left we have an instruction mix composed entirely of
  \texttt{vbroadcast} unit updates and on the right we have an instruction mix composed
  of \texttt{vshuffle} unit updates.}
\label{fig:tilings}
\end{figure}

Our complete work flow is captured in Figure~\ref{fig:workflow}, and
accomplishes the following: We start with the ISA for the target architecture,
this includes the available instructions and their latency and throughput. This
information is passed to our {\it Unit Update Enumeration} stage which
enumerates the space of all unit updates, for example given the ISA in
Figure~\ref{fig:simd_instruction_examples} the unit updates in
Figure~\ref{fig:components} are enumerated. After the unit update space is
enumerated, all possible tilings of unit updates that form outer-products of
our desired $m_r \times n_r$ are enumerated (Figure~\ref{fig:tilings}). These
outer-products are then modeled and estimated using our queueing
theory model described in the previous section. This process insures that the tiling, or
instruction mix, selected can sustain a high throughput, given that all other
instruction overheads are minimize. The steps that follow focus on minimizing
said overhead through several optimizations, namely static instruction
scheduling.

Continuing with the work flow, the highest performance instruction mix tiling is
selected and passed to a {\it kernel builder} which blocks the
outer-product. This will reduce register pressure when we perform instruction
scheduling (see Figure~\ref{fig:subblock_schedule}). The kernel builder
then outputs a skeleton of the kernel, like the one in
Figure~\ref{fig:code_example}, which captures the various blocking parameters
of the kernel along with  a set of {\it embedding functions} such as
Figure~\ref{fig:embedded_func}. These embedding functions capture the selected
instruction mix as functions. At this point the embedding functions and
skeleton represented an untuned matrix multiply kernel that is implemented
using the selected instruction mix.

The embedding functions and the skeleton are then passed to a {\it Scheduling and
  Optimization} phase which hides the instruction latency  by statically
performing software pipelining scheduling \cite{software_pipeline}, allowing
the kernel to perform near the predicted performance. The resulting statically
scheduled code is then emitted using inline assembly ANSI C intrinsics from
\cite{hands_off_hands_on}, which produces C code that can be compiled with a
fixed static schedule. This process yields outer-product kernel code
(Figure~\ref{fig:kernel_excerpt}) that achieves expert level performance. The
role of the external compiler on this emitted C code is to provide
register coloring, simplify memory indexing computation, and insure
efficient instruction alignments for the fetch and decode stages of the
processor.

\subsection{Embedding Function Capture the Instruction Mix}
The instruction-mix selected by our queuing model is not a complete kernel,
instead it is a collection of instructions that describe the data movement and
floating-point computation of data elements in a permuted outer-product. The
dependencies, register utilization and memory address computation is implicit
in this stage. Therefore, the first step in the transformation from
instruction-mix to kernel is to make these characteristics explicit. This is
done by expressing the three components of an outer-product instruction mix
(gathering of the elements $A$, gathering of the elements of $B$, performing
the multiply and accumulate of $C$) are expressed as embedding functions:
\texttt{get\_a\_element}, \texttt{get\_b\_element},  and \texttt{fma}
(Figure~\ref{fig:embedded_func}). In these functions dependencies,
register utilization, and memory address computation are made explicit.

\begin{figure}
  {\small
\begin{verbatim}
get_b_element( b_reg, ii, jj, pp )
 if( ii == 0 )
  switch( jj )
   case 0:
     b_reg[jj] = vload(&B[jj + pp*nr])
   case 1:
     b_reg[jj] = vshuffle(b_reg[jj-1])
   case 2:
     b_reg[jj] = vpermute(b_reg[jj-1])
   case 3:
     b_reg[jj] = vshuffle(b_reg[jj-1])

get_a_element( a_reg, ii, jj, pp )
 if( jj == 0 && ii mod v == 0)
  a_reg[ii] = vload( &A[ii + pp*mr] )

fma( a_reg, b_reg, c_reg, ii, jj, pp )
 if( ii mod v == 0 )
  c_reg[ii][jj] = vfma( a_reg[ii],
                        b_reg[jj],
                        c_reg[ii][jj] )
\end{verbatim}
}
  \caption{In order to pass the instruction mix to the kernel code generator, it is
    encoded as a function, similar to listing. These functions dispatch to a specific
    instruction which depends on what element $C_{ii,jj}$ is being on in the outer
    product.}
  \label{fig:embedded_func}
\end{figure}

These functions take the following inputs: arrays of registers that represent
the elements of $a$,$b$ and $c$, along with the indices for the $m$, $n$ and
$k$ dimension of the current unit-update. The embedding functions maps the
indices of the outer-product to the appropriate instructions from the
instruction-mix. Because these functions capture the majority of the work a
handful of optimizations are applied to these embedding functions. Namely, we
optimize for the fetch and decode stage by minimizing bytes per instruction,
and we simplify dependencies to minimize register utilization.

\paragraph{Optimizing for bytes per instruction.}
On most modern processors, the maximum throughput of the fetch and decode units
is low enough to become a bottleneck. Thus, some of the optimizations
performed by our generator minimize the instruction length and decode
complexity in order to avoid this bottleneck. The following decisions ensure
that shorter instructions are generated:

\begin{enumerate}
\item In some cases, we generate instructions that are meant to
  operate on single-precision data instead of instructions that
  operate on double-precision data. An example of this is the use of
  the \texttt{vmovaps} instruction to load \texttt{reg\_a}, instead of
  \texttt{vmovapd}. This is because both instructions perform the identical
  operation but the single-precision instruction can be encoded in fewer bytes.

\item We hold the partially accumulated intermediate $m_r \times n_r$
  matrix of $C$, which we will refer to as $T$, using high-ordered
  registers (i.e. register \texttt{xmm8} to \texttt{xmm16}). On most
  architecture we tested, high-ordered SIMD registers require more
  bytes to encode. Thus, by using the low-ordered registers to hold
  working values and high-ordered registers to store $T$, we ensure
  that each instruction has at most one register operand (i.e. the
  output operand) that is a high-ordered register.

\item For memory operations, address offsets that are beyond the range of
  $-128$ to $127$ bytes  require additional bytes to encode. Therefore, we
  restrict address offsets to fit in this range by subtracting $128$ bytes
  from the base pointers into $A$ and  $B$.
\end{enumerate}

\paragraph{Eliminating unnecessary dependencies.}
 Notice that each permute-multiply-add step of a unit update can be  performed
 independently  once the permutation of \texttt{reg\_b} has been
 completed. The key is to ensure that in generating the permutations,
 unnecessary  dependencies  are not introduced to make the independent
 permute-multiply-add steps  dependent. Consider the following code snippet
 for producing four permutations of $B$:
\begin{center}
\begin{verbatim}
  vmovapd    (addr_B), %ymm1
  vshufpd    $5,       %ymm1, %ymm1, %ymm1
  vperm2f128 $1,       %ymm1, %ymm1, %ymm1
  vshufpd    $5,       %ymm1, %ymm1, %ymm1
\end{verbatim}
\end{center}
  Notice that each instruction is dependent on the result of the
  permutation instruction immediately before it. As a result, the previous set of
  instructions would take longer to compute than the following sequence of  instructions.
\begin{center}
\begin{verbatim}
  vmovapd    (addr_B), %ymm1
  vperm2f128 $1,       %ymm1, %ymm1, %ymm2
  vshufpd    $5,       %ymm1, %ymm1, %ymm3
  vshufpd    $5,       %ymm2, %ymm2, %ymm4
\end{verbatim}
\end{center}
 In both cases the same permutations are being computed, but in the latter
 case the permutations are stored in different registers. This eliminates
 the false dependencies between the instructions, so the shuffle
 instructions are independent and can be executed independently.

\subsection{From Embedding Functions to Generated Code}
\begin{figure} 
\includegraphics[scale=.4]{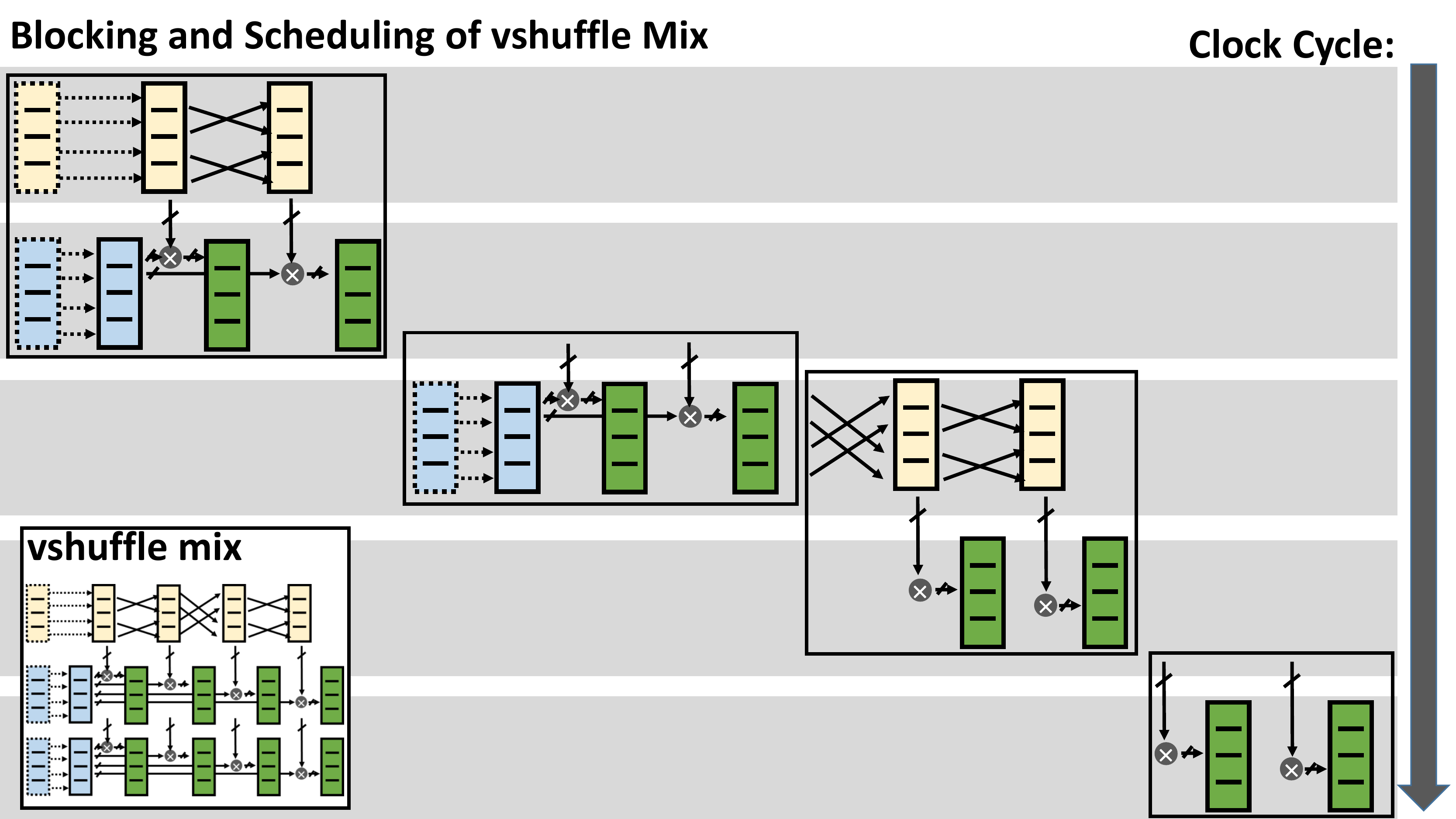}
  \caption{Once a candidate outer product tiling is selected
    (figure~\ref{fig:tilings}), we perform an additional layer of blocking
    ($m_s$, and $n_s$) to assist the code generator in minimizing register
    spills. Register blocking allows fewer registers to be live at a given
    cycle thus allowing the code generator to aggressively schedule the
    instructions.}
  \label{fig:subblock_schedule}
\end{figure}

Once we have an instruction mix for the outer product, as determined by our
queueuing theory model, we can generate an implementation that hides the
instruction latency in this instruction mix. Static instruction scheduling is
key for this next step, however this optimization is limited by the number
of available registers. Thus, the primary step in the {\it kernel builder}
(Figure~\ref{fig:workflow}) is to determine a further layer of blocking for
the outer product.

\begin{figure} 
  {\small
\begin{verbatim}
/* initialize temp buffer */
for( i = 0; i <  m_r; i++ )
 for( j = 0; j <  n_r; j++ )
  c_reg[ii][jj] = 0;

/* computation */
#unroll(k_u)
#schedule_software_pipeline
for( pp = 0; pp < k_b; pp++ )
 /* perform the outer products */
 for( i = 0; i <  m_r; i+=m_s )
  for( j = 0; j <  n_r; j+=n_s )
   for( ii = i; ii < i+m_s; ii++ )
    get_a_elem(a_reg, ii,j );
    for( jj = j; jj < j+n_s; jj++ )
     get_b_elem(b_reg, ii,jj );
     fma( c_reg, a_reg, b_reg, ii,jj,pp );

/* accumulate temp to results */
for( i = 0; i <  m_r; i++ )
 for( j = 0; j <  n_r; j++ )
    C[(ii,jj)] += c_reg[ii][jj];
\end{verbatim}
}
  \caption{In this code skeleton we capture an outline of the generated kernel. We pass a
    similar outline to our code generated along with our instruction mix. This
    mix is encoded as the functions \texttt{get\_a\_elem},
    \texttt{get\_b\_elem} and \texttt{fma}. The code generator uses this
    information to generate the code, perform optimizations such as unrolling
    and code motion and schedule the resulting kernel code using software
    pipelining in way that targets the microarchitecture.}
\label{fig:code_example}
\end{figure}

\subsection{Limits imposed by registers.}
Recall that to compute a unit update, $u_b$ permutations of the elements in
\texttt{reg\_b} are required. However, a multiply and an add is performed with
each permutation. This implies that each unit update will require two
new registers ($R_{R}=2$): One to store the permutation of \texttt{reg\_b},
and another to hold the output of the multiplication.  On architectures with a
{\em fused-multiply-add} instruction, only one new register is required (i.e. $R_{R}=1$).

As the register that holds the output of the accumulated result is reused over
multiple outer-products, it means that the number of unit updates
($N_{\mbox{\scriptsize {updates}}}$)  that can be performed without register
spilling is constraint only by the number of registers given by the following:

\begin{equation}
N_{\mbox{\scriptsize {updates}}} = \left\lfloor \frac{R_{total} - \frac{m_r
    n_r}{v} - R_{A}}{R_{R}}\right\rfloor,
\label{eqn:nupdate}
\end{equation}
where $R_{total}$ and $R_A$ are the total number of registers and the
registers required to hold the the column of $A$. We select the additional
blocking dimensions such that:
\begin{equation}
m_s n_s \le N_{\mbox{\scriptsize {updates}}} v
\end{equation}
In Figure~\ref{fig:subblock_schedule} we show the blocking and scheduling of a
\texttt{vshuffle} instruction mix (Figure~\ref{fig:tilings}) for a $m_r \times
n_r = 8 \times 4$ outer-product.

\subsection{Scheduling and Tuning}

The instruction mix selected by the {\it Queueing Model Estimator} is
translated by the {\it Kernel Builder}  into several {\it embedding functions}
(\texttt{get\_a\_elem}, \texttt{get\_b\_elem} and \texttt{fma}), like those in
Figure~\ref{fig:embedded_func}. These function are embedded in a looping structure
that matches the outer-product kernel (Figure~\ref{fig:code_example}). These loops
iterate over the $ m_r, n_r, m_s $ and $n_s$ dimensions and generate the
dependencies between the instructions inside the embedding functions.

Once
these dependencies are built, a few basic optimizations, such as common
sub-expression elimination, are performed such that the only the original
instruction mix plus a few looping instructions exist in the final output
code.

Next, the code generator performs software pipelining
\cite{software_pipeline} over the entire looping structure of the
outer-product. By statically scheduling the kernel, the risk of instruction
stalls is minimized thus allowing the processor to compute the instruction mix
near the rate predicted by our model. Once the code is scheduled, the
generator emits a mix of C code and inline assembly instruction macros which
preserve the schedule \cite{hands_off_hands_on}. This resulting code
implements a high performance  outer-product kernel with the desired
dimensions and the selected instruction mix. We provide an excerpt of a
generated kernel in Figure~\ref{fig:kernel_excerpt}.

\begin{figure} 
  {\small
\begin{verbatim}
for( pp = 0; pp < k_b; pp+=KUNR )
{
 /* STEADY STATE CODE */
 VLOAD_IA(GET_A_ADDR(0),GET_A_REG(0))
 VLOAD_IA(GET_A_ADDR(1),GET_A_REG(1))
 VLOAD_IA(GET_B_ADDR(0),GET_B_REG(0))
 VSHUFFLE_IA(0x05,GET_B_REG(0),GET_B_REG(1))
 VFMA(GET_A_REG(0),GET_B_REG(0),GET_C_REG(0,0))
 VFMA(GET_A_REG(0),GET_B_REG(1),GET_C_REG(0,1))
 VPERM2F128_IA(0x01,GET_B_REG(1),GET_B_REG(2))
 VSHUFFLE_IA(0x05,GET_B_REG(2),GET_B_REG(3))
 VFMA(GET_A_REG(1),GET_B_REG(0),GET_C_REG(0,0))
 VFMA(GET_A_REG(1),GET_B_REG(1),GET_C_REG(0,1))
 /* snip */
}
\end{verbatim}
}
  \caption{This is generated excerpt from our kernel generator. The resulting
    kernel code implements the instruction mix identified by our queueing
    theory model and is statically scheduled to maintain the estimated
    performance of the mix.}
\label{fig:kernel_excerpt}
\end{figure}

\section{Experimental Results}

In this section, we test the effectiveness of our kernel generation system in
automating the last-mile for high performance dense linear algebra. We
evaluate both the queueing theory model, which finds an efficient
outer-product instruction-mix, and the code generation system, which
translates that mix into a high performance kernel.

We use a variety of machines listed in Table~\ref{tab:machine_table} that span
a diverse range of double precision vector lengths ($v\in \{2,4,8\}$), number
and partitioning of functional units, and instruction latencies. Because our
kernel fits in the context of a larger Goto/BLIS-style \Gemm context, the
blocking parameters $m_c,k_c,m_r$ and $n_r$ are determined from the analytical
models developed in \cite{Goto:2008:AHP} and \cite{BLIS4} along with the cache
and microarchitecture details listed in Table~\ref{machine_table_cache} and
Table~\ref{machine_table_uarch} respectively. The microarchitecture details in
particular (Table~\ref{machine_table_uarch}) were used by the queueing theory
model to select the highest throughput outer-product
instruction-mix. Additionally, these details determined $N_{\mbox{\scriptsize {updates}}}$ and the register
sub-blocking dimensions $m_s,n_s$ using the formula developed in the previous
section.

Lastly, the Xeon Phi requires that four threads run concurrently in order to
effectively utilize a core. This requires that we distribute the work across
multiple threads. Therefore we used  the implementation in
\cite{BLIS3} with the following parameters:  the number of threads used in
each dimension ($i_c$ and $j_r$) must satisfy  $i_c * j_r \le 59*4$, and
ideally should be factors of  $\frac{m}{m_c}$ and $n$ respectively. By
empirical selection, $i_c=12$ and $j_r=16$ satisfied both of those
requirements and resulted in the largest number of cores that achieved
efficient per core performance.

\begin{landscape}
\begin{figure}
  {\small
    \begin{tabular}{|l||r|r|r|r|r|r|r|r|r|r|r|}
      \hline
      Proc.  &
      uArch. &
      Freq.  &
      $S_{\textrm{L1}}$ &
      $W_{\textrm{L1}}$ &
      $N_{\textrm{L1}}$ &
      $S_{\textrm{L2}}$ &
      $W_{\textrm{L2}}$ &
      $N_{\textrm{L2}}$ &
      $S_{\textrm{L3}}$ &
      $W_{\textrm{L3}}$ &
      $N_{\textrm{L3}}$ \\
      \hline
      \hline
      Core 2 X9650   &
      Penryn &
      3 GHz  &
      $4 \times 32$ KiB &
      8 &
      256 &
      $2 \times 6$  MiB&
      24 &
      16384 &
      - &
      - &
      - \\
      \hline
      Xeon X5680  &
      Nehalem &
      3.333 GHz  &
      $6 \times 32$ KiB&
      $8$ &
      $64$ &
      $6 \times 256$ KiB&
      $8$ &
      $512$ &
      $ 12 $ MiB&
      $16$ &
      $12288$ \\
      \hline
      Core i5-2500  &
      Sandy Bridge &
      3.3 GHz  &
      $4 \times 32$ KiB &
      4 &
      512 &
      $4 \times 256$ KiB &
      4 &
      4096 &
      6 MiB &
      12 &
      32768 \\
      \hline
      Core i7-4770K &
      Haswell &
      3.5 GHz  &
      $4 \times 32$ KiB &
      8 &
      256 &
      $4 \times 256$ KiB &
      8 &
      2048 &
      8 MiB &
      16 &
      32768 \\
      \hline
      Xeon Phi 5110p  &
      Knights Corner &
      1.053 GHz  &
      $60 \times 32$ KiB &
      8 &
      256 &
      $60 \times 512$ KiB &
      8 &
      4096 &
      - &
      - &
      - \\
      \hline
    \end{tabular}
  }
  \caption{ Cache details of the processors used in our experiments. These
    cache details are needed for determining $m_c,k_c,m_r$ and $n_r$ according
    to \cite{Goto:2008:AHP} and \cite{BLIS4}. The value $S_l$ corresponds to
    the size of the $l$ level of cache. $W_l$ is the number of ways and $N_l$
    is the number of cache lines in each way.}
  \label{machine_table_cache}
\end{figure}

\begin{figure}[h!]
  {\small
    \begin{tabular}{|l||r|r|r|r|r|r|r|r|r|r|r|r|}
      \hline
      uArch.  &
      Reg. &
      $\ell_{\textrm{fma}}$ &
      $\ell_{\textrm{L1}}$ &
      $\ell_{\textrm{L2}}$ &
      $\ell_{\textrm{shuf.}}$ &
      $\ell_{\textrm{perm.}}$ &
      $\ell_{\textrm{bcast.}}$ &
      $R_{\textrm{fma}}$ &
      $R_{\textrm{mem}}$ &
      $R_{\textrm{shuf.}}$ &
      $R_{\textrm{perm.}}$ &
      $R_{\textrm{bcast.}}$ \\
      \hline
      \hline
      Penryn & 
      16 &
      $5 + 3$ &
      $3$ &
      $15$ &
      $1$ &
      - &
      $1$ &
      $p_0 \wedge p_1$ &
      $p_2$ &
      $p_5$ &
      - &
      $p_0$ \\
      \hline
      Nehalem & 
      16 &
      $5 + 3$ &
      4 &
      10 &
      $1$ &
      - &
      2 &
      $p_0 \wedge p_1$ &
      $p_2$ &
      $p_0 \vee p_5$ &
      - &
      $p_5$ \\
      \hline
      Sandy Bridge & 
      16 &
      $5 + 3$ &
      4 &
      12 &
      1 &
      2 &
      3 &
      $p_0 \wedge p_1$ &
      $p_2 \vee p_3$ &
      $p_5$ &
      $p_5$ &
      $p_5 \wedge (p_2 \vee p_3)$ \\
      \hline
      Haswell & 
      16 &
      $5$ &
      $4$ &
      $12$ &
      $1$ &
      $3$ &
      $5$ &
      $p_0 \vee p_1$ &
      $p_2 \vee p_3$ &
      $p_5$ &
      $p_5$ &
      $p_2 \vee p_3$ \\
      \hline
      \hline
      uArch.  &
      Reg. &
      $\ell_{\textrm{fma}}$ &
      $\ell_{\textrm{L1}}$ &
      $\ell_{\textrm{L2}}$ &
      $\ell_{\textrm{shuf. fma}}$ &
      $\ell_{\textrm{perm.}}$ &
      $\ell_{\textrm{bcast. fma}}$ &
      $R_{\textrm{fma}}$ &
      $R_{\textrm{mem}}$ &
      $R_{\textrm{shuf. fma}}$ &
      $R_{\textrm{perm.}}$ &
      $R_{\textrm{bcast. fma}}$ \\
      \hline
      \hline
      Knights Corner & 
      32 &
      $4$ &
      $1$ &
      $11$ &
      $4$ &
      $6$ &
      $4$ &
      $p_0$ &
      $p_{\textrm{mem}}$ &
      $p_0 \wedge p_{\textrm{mem}}$ &
      $p_0$ &
      $p_0 \wedge p_{\textrm{mem}}$ \\
      \hline

  \end{tabular}
  }
  \caption{Here we capture the pertinent microarchitecture parameters that are
    used for our queueing theory model. The column $\ell_u$ represents the
    latency in cycles of instruction $u$, where L1 and L2 represents
    instruction reads that hit in those caches. In the case of a system
    without fused-multiply-add (fma), the latency is represented as the sum of
    the multiply instruction and add instruction. The columns $R_u$ represent
    the functional units that are required to compute instruction $u$. For
    some instructions multiple function units may be required (represented by
    $\wedge$) and some instruction may take multiple paths (represented by $\vee$).}
  \label{machine_table_uarch}
\end{figure}


\begin{figure}[h!]
{\small
  \begin{tabular}{|l||r|r|r|r|} 
    \hline
        {\bf Processor}( {\bf uArch.})
        & ${ m_c \times k_c} $
        &${  m_r \times n_r}$
        &$N_{\mbox{\scriptsize {updates}}}$
        &${ m_s \times n_s}$\\ \hline
        \hline
    Core 2  X9650 (Penryn) & $256 \times 256$ & $4 \times 4$ & 3 &$2 \times 2$   \\ \hline
    Xeon X5680 (Nehalem) & $256 \times 256$ & $2 \times 8$ & 3 &$2 \times 2$ \\ \hline
    Core i5-2500 (Sandy Bridge) & $96 \times 256$ & $8 \times 4$ & 3 &$4 \times 2$  \\ \hline
    Core i7-4770K (Haswell) & $256 \times 512$ & $4 \times 12$ & 3 &$4 \times 4$ \\ \hline
    Xeon Phi 5110p (Xeon Phi) & $120 \times 240$ & $30 \times 8$ &1 &  $8 \times 1$ \\
    \hline
  \end{tabular}
  }
\caption{ The cache blocking parameters $m_c$ and $k_c$ where determined using the results in
  \cite{Goto:2008:AHP} and the hardware parameters in
  Table~\ref{machine_table_cache} and Table~\ref{machine_table_uarch}. The
  register blocking parameters $m_r$ and $n_r$ were determined from
  \cite{BLIS4} using the values in Table~\ref{machine_table_cache}. Lastly,
  $N_{\mbox{\scriptsize {updates}}}$ and subsequently the sub-blocking
  dimension $m_s$ and $n_s$ were determined using
  Equation~\ref{eqn:nupdate}. $m_c,k_c,m_r,n_r,m_s $ and $n_s$ correspond to
  the values used in the   generated code (see
  Figure~\ref{fig:code_example}). Note $N_{\mbox{\scriptsize {updates}}}v \ge
  m_sn_s$}.
\label{tab:machine_table}
\end{figure}
\end{landscape}

\subsection{Analysis of Queueing Model}

\begin{figure}[h!]
  \center
{\small
  \begin{tabular}{ | l | l |l |l | l| l| l | } 
    \hline
\multicolumn{7}{|c|}{ {\bf Xeon Phi $m_r\times n_r=8\times 30$ Impl.} }\\ \hline
{\bf \shortstack{\# vperm.\\ updates}}&	
{\bf \shortstack{\# vbcast.\\ updates}}&	
{\bf \shortstack{\# Reads \\($L_{\textrm{mem}}$)} } &
{\bf \shortstack{\# Vect. \\($L_{p_0}$)}}&
{\bf
  \shortstack{$\lambda_{\textrm{outer-product}}$\\ $\min(\frac{1}{L_{\textrm{mem}}},\frac{1}{L_{p_0}})$}} &
{$\frac{\textrm{flop}}{\textrm{cyc.}}$} &
Est. $\frac{\textrm{GFLOP}}{s}$ \\

\hline
\hline
0&	30&	1+0+30+4=35& 31&	0.02857 & 13.71& 14.44 \\    \hline
1&	26&	1+1+26+4=32& 32&	0.03125 & 15   & 15.80\\    \hline
2&	22&	1+2+22+4=29& 33&	0.03030 & 14.55& 15.32\\    \hline
3&	18&	1+3+18+4=26& 34&	0.02941 & 14.12& 14.87\\    \hline
4&	14&	1+4+14+4=23& 35&	0.02857 & 13.71& 14.41\\    \hline
5&	10&	1+5+10+4=20& 36&	0.02778 & 13.33& 14.04\\    \hline
6&	6&	1+6+6+4 =17& 37&	0.02703 & 12.97& 13.66\\    \hline
7&	2&	1+7+2+4 =14& 38&	0.02632 & 12.63& 13.30\\    \hline
  \end{tabular}
}
  \caption{ We estimate the  number of cycles needed to compute
    our generated Xeon Phi outer-product kernels. The first column is the
    number of \texttt{vpermute} unit updates of size $4\times 8$ used to
    implement the outer product. The remainder of the outer-product is
    computed using multiple $1\times 8$  broadcast based unit updates.}
\label{tab:mic_prediction}
\end{figure}



%

In order to demonstrate the effectiveness of our model, we compare the
predicted performance against the actual performance estimated by our queueing
theory model.

\begin{figure*}
\centering
\begin{tabular}{l}

  \includegraphics[clip, trim=1cm 2cm 1cm 2cm,scale=.5]{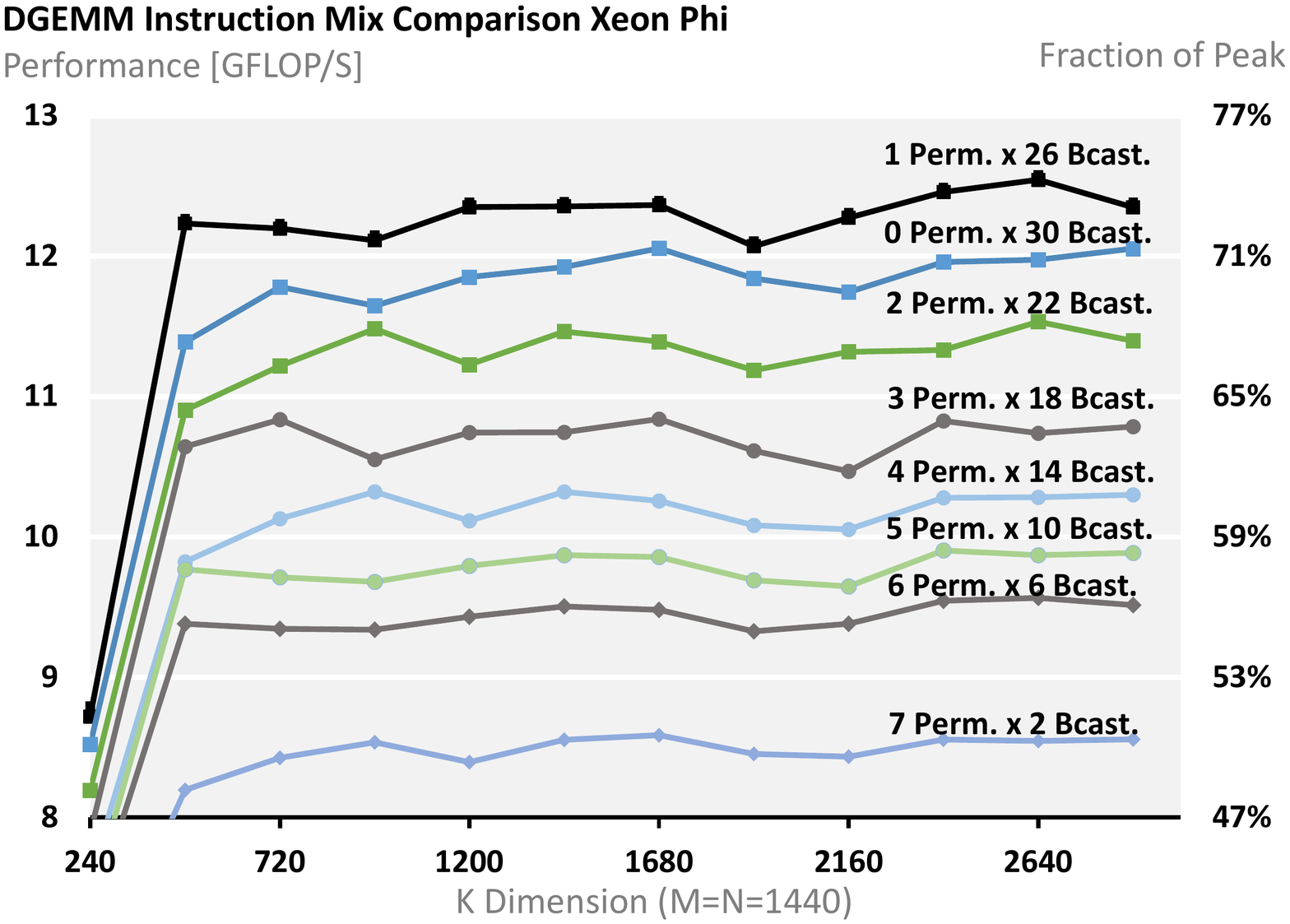}
\\
  \includegraphics[clip, trim=1cm 2cm 1cm 2cm,scale=.5]{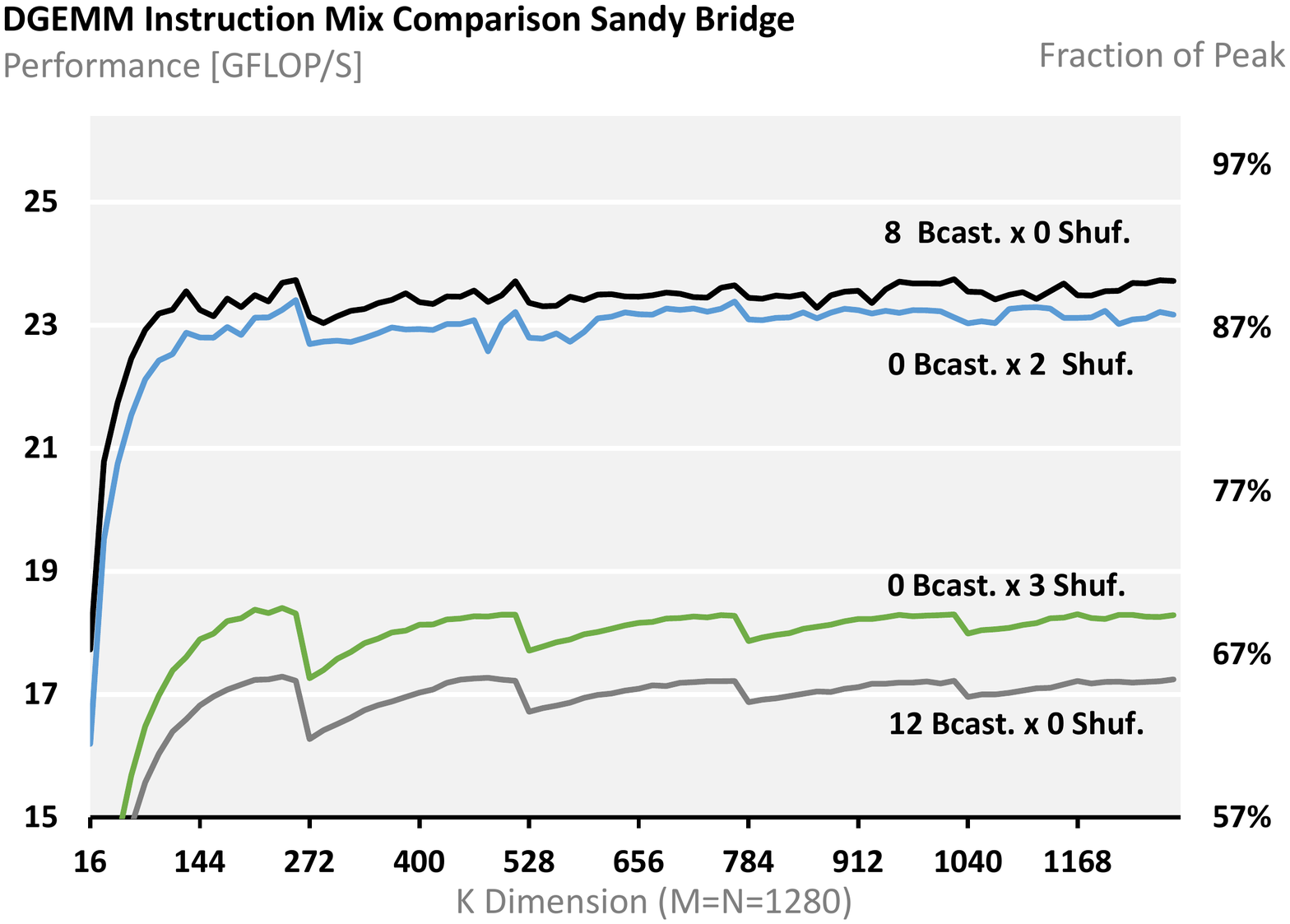}
\\
\end{tabular}
\caption{In both of these experiments we test the accuracy of our queueing
  theory model. {\bf Top:} On the Xeon Phi we evaluate the performance of
  eight implementations of the same outer-product. {\bf Bottom:} We do a
  similar experiment, but with two different outer-product sizes.}
\label{fig:model_evaluation}
\end{figure*}

For the Xeon Phi we compare the performance of eight different instruction-mix
implementations of an $8\times 30$ outer-product. We selected a family of
instruction-mixes where the work is partitioned between $8 \time 4$ permute unit
updates and $8 \times 1$ broadcast based unit updates. In
Table~\ref{tab:mic_prediction} we detail each outer-product
implementation. Each row represents a specific implementation, where the first
two columns represent the number of permute and broadcast unit updates in the
implementation.In the next two columns we compute the number of instructions
that need the memory port ($p_{\textrm{mem}}$) and vector port ($p_0$)
respectively. For the Xeon Phi each permute component requires one load
instruction, a permute instruction and four fma instructions. The broadcast
based component require one load and one fma instruction. Additionally, each
implementation requires four prefetch instructions that occupy the memory
ports. In the fifth column we use our queueing theory model to estimate the
performance of the implementation. We can estimate the performance in FLOP per
cycle as:
\begin{equation}
  \frac{\textrm{flop}}{\textrm{cyc.}} = \lambda_{\textrm{outer-product}}  (m_r)(n_r)(2 flop)
\end{equation}
In the last column we estimate the performance in GFLOP using the following
formula:
\begin{equation}
  \frac{\textrm{GFLOP}}{\textrm{s}} = f\frac{\textrm{flop}}{\textrm{cyc.}}
\end{equation}
Each of these implementations has a different throughput predicted by our
model. In our experiment (Figure~\ref{fig:model_evaluation}), we compare the
relative performance of these implementations. Assuming that the overheads are
similar between all implementations, then if the model does not fit, we expect
a significant difference between the relative ordering of the implementations
and the predictions. However, for the Xeon Phi we see that the relative
ordering of the implementations is preserved in the experimental results, with
the exception of one of the implementations. We suspect that the overhead is
slightly lower for the {\it 0 permute, 30 broadcast} implementation.

\begin{figure}[h!]
  \center
{\small
  \begin{tabular}{ | l | l |l | l| l| l | l| l| l| } 
    \hline
    $m_r \times n_r$ &
    {\bf \shortstack{\#bcast.\\ updt.}}&
    {\bf \shortstack{\#shuf.\\ updt.}}&
    {\bf \shortstack{\#mem.\\ $L_{p_2 \vee p_3}$}}&
    {\bf \shortstack{\#fma\\ $L_{p_0 \wedge p_1}$}}&
    {\bf \shortstack{\#shuf. \\$L_{p_5}$} } &
    {\bf   $\lambda_{\textrm{out.-prod.}}$ } &
    {$\frac{\textrm{flop}}{\textrm{cyc.}}$} &
    $\frac{\textrm{GFLOP}}{s}$ \\
\hline
\hline
$8 \times 4$  & 8  & 0 & 2 +  8 = 10 & 8  & 8  & 0.125 & 4 & 26.4\\ \hline
$8 \times 4$  & 0  & 2 & 2 +  1 = 3  & 8  & 3  & 0.125 & 4 & 26.4\\ \hline
$4 \times 12$ & 0  & 3 & 1 +  3 = 4  & 12 & 9  & 0.083 & 4 & 26.4\\ \hline
$4 \times 12$ & 12 & 0 & 1 + 12 = 13 & 12 & 12 & 0.083 & 4 & 26.4\\ \hline
  \end{tabular}
}
  \caption{ We estimate the  number of cycles needed to compute
    our generated Sandy Bridge outer-product kernels. We implement
    outer-products of size $m_r \times n_r \in \{8 \times 4, 4 \times
    12\}$. Note that the model predicts similar performance across these
    implementation, however due to subtle microarchitecture details the
    experimental performance is different.
  }
\label{tab:snb_prediction}
\end{figure}

We repeat the same experiment with the Sandy Bridge processor. In
Table~\ref{tab:snb_prediction} we estimate the performance for several
implementations. Unlike the Xeon Phi experiment we chose two different kernel
sizes ($m_r \times n_r$). According to \cite{BLIS4}, the $8 \times 4$
implementations is more efficient than the $4 \times 12$. In
Figure~\ref{fig:model_evaluation} we plot the performance of the four
implementations. What we see is despite that our model predicts identical
performance, we see a significant difference between the kernels of different
sizes. What this demonstrates is that even if we can produce an
efficient kernel in isolation, our model operates within the
constraints of the larger GotoBLAS/BLIS algorithm. There are also additional
and subtle microarchitectural details that explain the difference between
implementations of the same size on this system. For example, even though both
ports $p_2$ and $p_3$ service memory operations, they are limited in the total
number of bytes that can be read in a cycle. Therefore, the Sandy Bridge
retires less than 2 memory operations per cycle. In the case where this is not
an issue (between the two $8 \times 4$ implementations, we attribute the
performance difference to scheduling because the permute based approach has
fewer dependencies than the broadcast implementation, giving the scheduler
greater freedom to hide instruction latency.

This experiment demonstrates that for outer-product implementations of same
size we can accurately estimate the performance of our generated
implementations. However, our kernels operate within the constraints of a
bigger GotoBLAS/BLIS \Gemm algorithm, and our performance is ultimately
limited by the parameters selected for the bigger algorithm. In the next
subsections, we look at this interaction in the opposite direction, how
decisions made in generating the kernel affect the overall GotoBLAS/BLIS
algorithm.

\subsection{Analysis of the Generated Kernel}

\begin{figure*}
\centering
\begin{tabular}{lr}
\includegraphics[clip, trim=1cm 2cm 1cm 2cm,scale=.5]{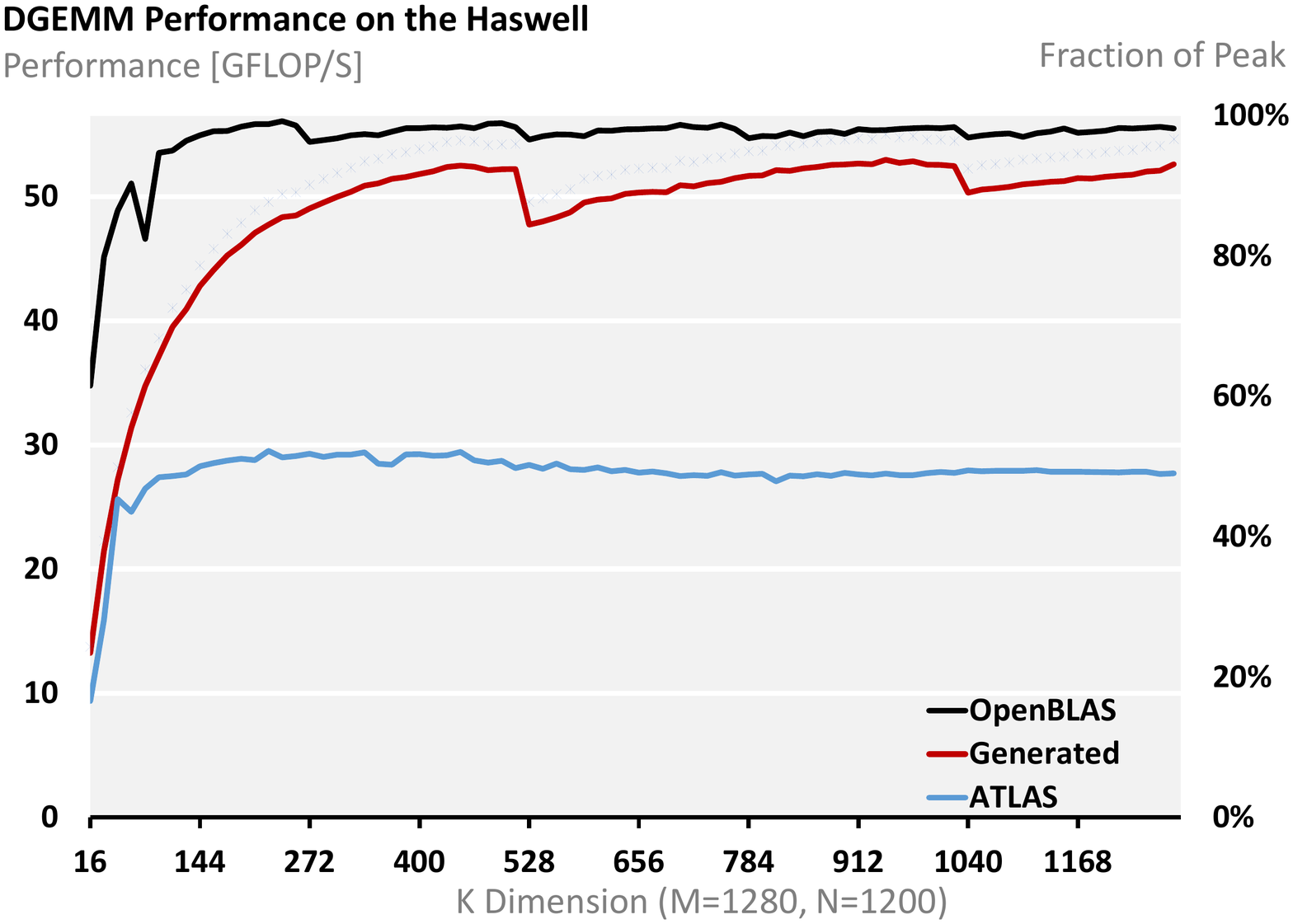}
\\
  \includegraphics[clip, trim=1cm 2cm 1cm 2cm,scale=.5]{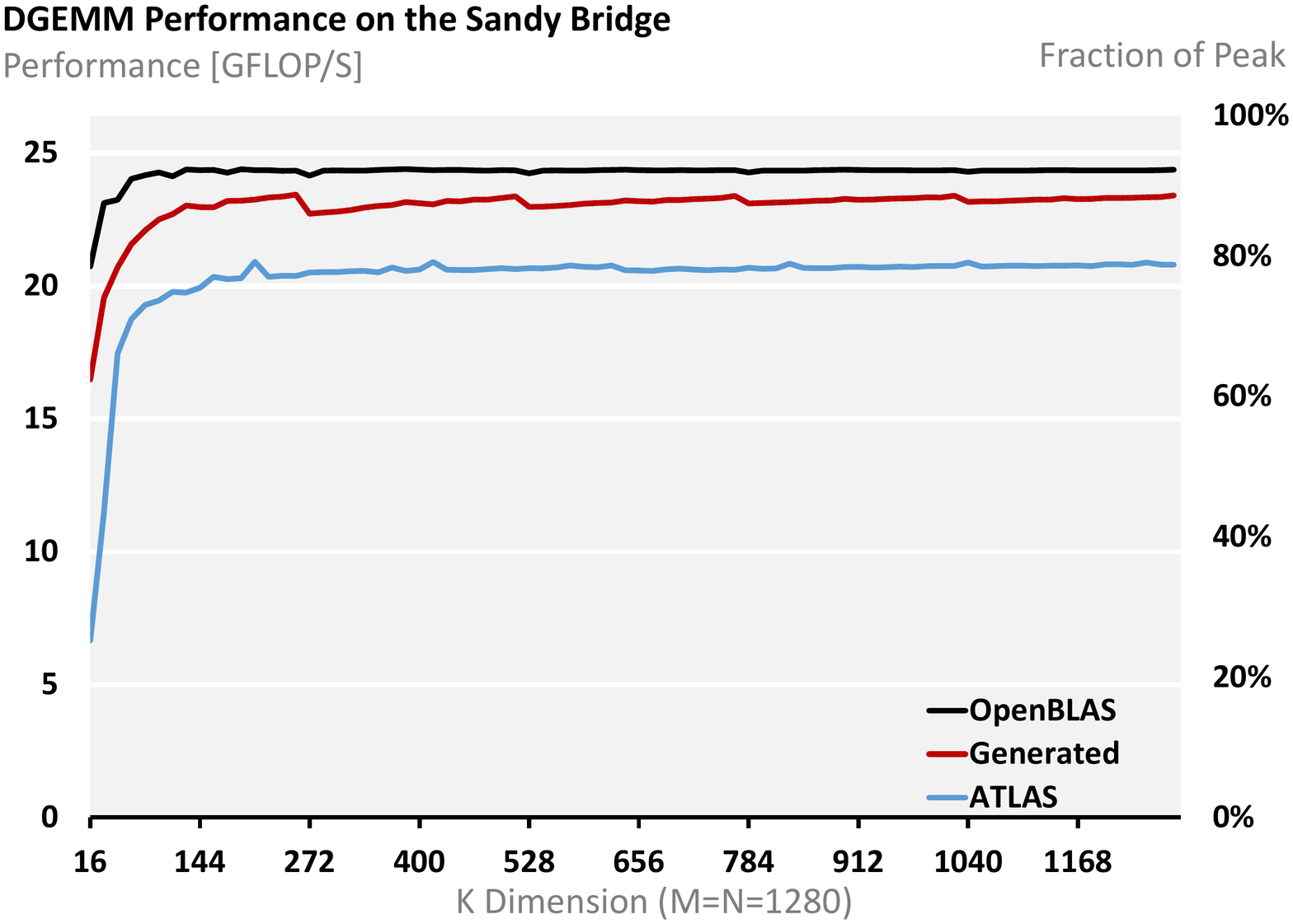}
\\
\end{tabular}
\caption{ We compare the performance of our generated kernels against ATLAS
  and the  OpenBLAS for various problem sizes to demonstrate that expert level
  performance can be automated. We see that our generated code approaches the performance of hand tuned expert code and for
  most architectures exceeds the performance of the generated ATLAS code.}
\label{all_sequential}
\end{figure*}

\begin{figure*}
\centering
\begin{tabular}{l}
\includegraphics[clip, trim=1cm 2cm 1cm 2cm,scale=.5]{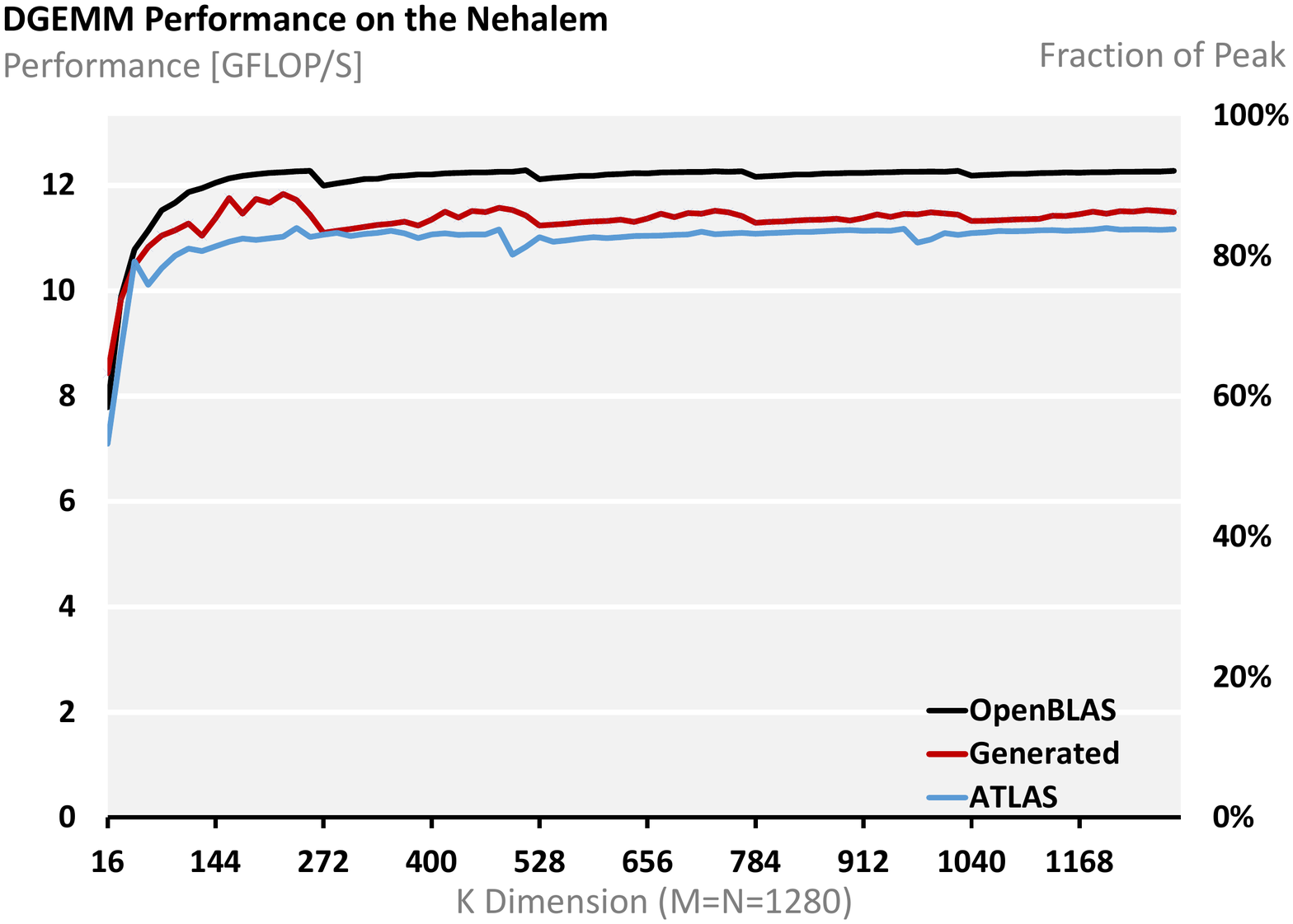}
\\
\includegraphics[clip, trim=1cm 2cm 1cm 2cm,scale=.5]{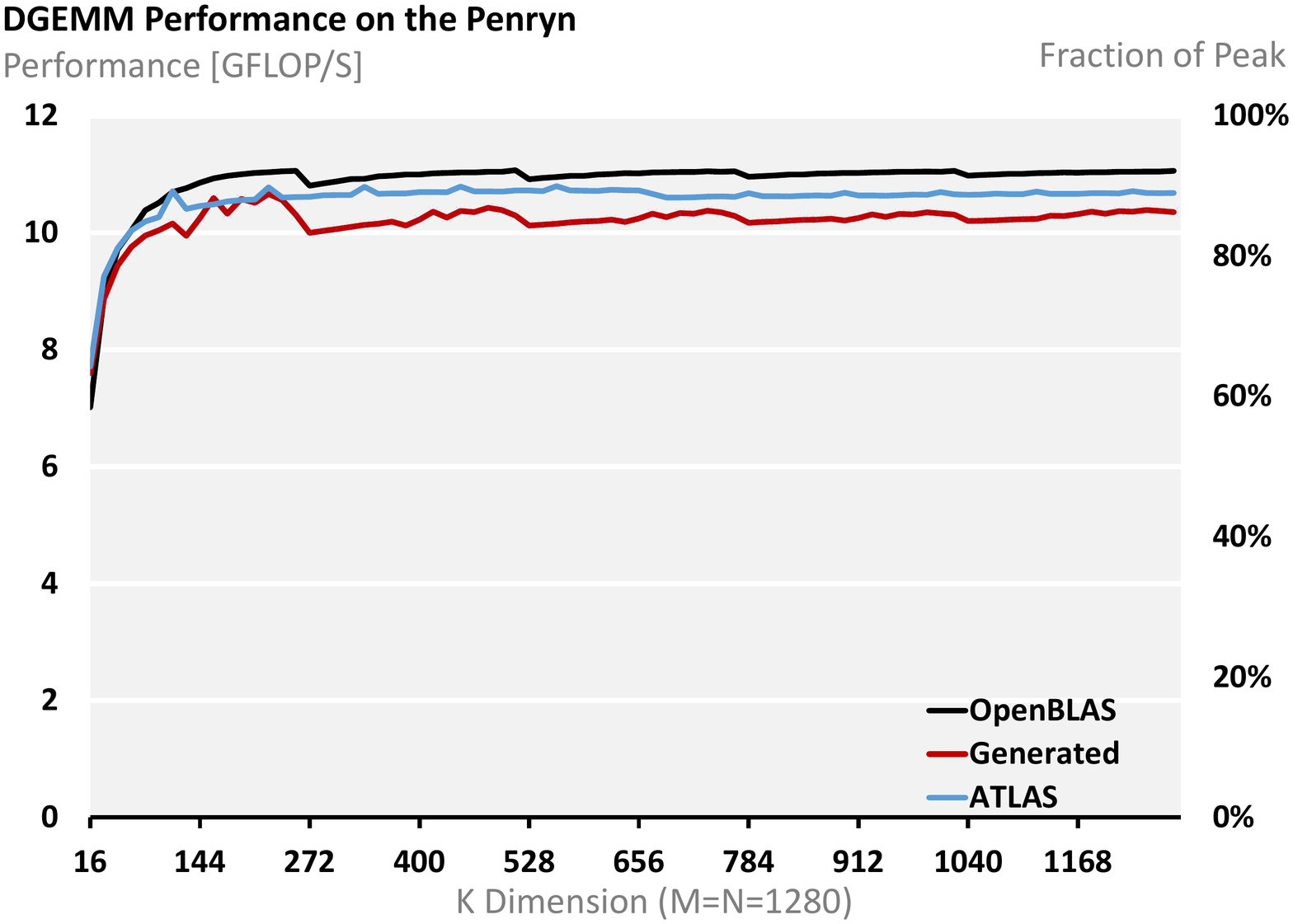}
\\
\end{tabular}
\caption{ Like the graphs in Figure~\ref{all_sequential}, we compare the
  performance of our generated kernels against the OpenBLAS and ATLAS. }
\label{all_sequential_p2}
\end{figure*}

%
%
%

We evaluate the effectiveness of our kernel generation approach by comparing
the performance of our generated outer-product kernels against state-of-the-art
{\Gemm} implementations such as OpenBLAS \cite{OpenBLAS} and ATLAS \cite{ATLAS}. We
selected OpenBLAS  because it is the highest performance open source BLAS
implementation on most architectures, including the systems used in this
paper. ATLAS was also selected because it is a high performance code
generation system. Unlike, our code generator, this framework relies on hand
tuned assembly kernels and uses search to determine the blocking dimensions
around these kernels. The systems used in this experiment represent the past
four major microarchitecture designs from Intel
Table~\ref{tab:machine_table}. The parameters for the GotoBLAS/BLIS \Gemm
algorithm were analytically selected to maximize performance. These values
also match the ones used by the OpenBLAS.

In these experiments (Figure~\ref{all_sequential} and
Figure~\ref{all_sequential_p2}), our generated code is within 2-5\% of the
expert tuned OpenBLAS. We suspect this difference is due to loop overhead
because we rely on the compiler to optimize this which results in several
extra instructions over the expert code. The older the processor generation,
the more pronounced of an effect this has, which is why ATLAS outperforms our
code on the Penryn. We believe we can resolve this difference by implementing
the looping structure in inline assembly which should give us performance that
is near identical to the expert written code.

\subsection{Sensitivity to Parameters}

\begin{figure*}
  \includegraphics[clip, trim=1cm 5cm 1cm 4cm,scale=.4]{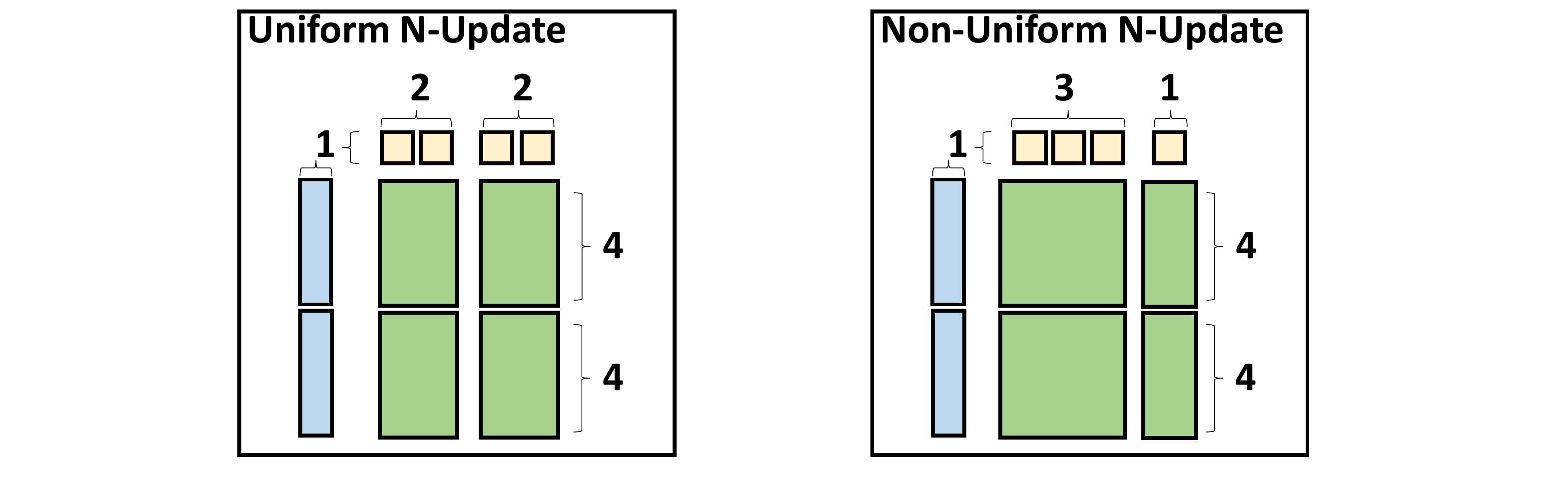}
  \caption{Given an outer-product instruction-mix of an $m_r \times n_r =
    8\times 4 $, we can partition it into uniformly sized N-updates or
    non-uniform size N-updates. In the uniform case, each N-update is
    $m_s\times n_s = 4 \times 2$, in the non-uniform case each N-update is
    $m_s\times n_s \in \{4 \times 3, 4 \times 1\}$. }
  \label{fig:nupdate_block_diagram}
\end{figure*}

\begin{figure}
\begin{center}
  \begin{tabular}{ |l|| l| l| l| l| l| l|  }
\hline
      Port ($R_U$)          &  $p_0$   &  $p_1$   &  $p_2$  -  $\ell_D$    &  $p_3$   -  $\ell_D$      &  $p_5$   \\ \hline
      Cycles - Uniform & 32.0 & 32.0 & 8.0 -   12.0 & 8.0  -  16.0  &
      16.0\\ \hline
     Cycles - Non-Uniform & 32.0 & 32.0 & 8.0 -   12.0 & 8.0  -  16.0  & 16.0 \\
\hline
  \end{tabular}
\end{center}
\caption{IACA results comparing instruction throughput between the uniform and
  non-uniform N-update implementations. Each port represents a functional
  unit that is used for our operation. $\ell_D$ represents data fetch
  latency. What this shows is that both the uniform and non-uniform
  shaped implementations of the same outer product look identical to the
  Out-of-Order engine, as simulated by the IACA tool. However, we will show
  the performance of the two implementations are significantly different. }
\label{iaca_n_update}
\end{figure}

In addition to using static scheduling and avoiding register spilling, we
observe that even in the presence of an Out-Of-Order engine there is a benefit
from maintaining uniformly-sized N-updates for creating the outer
product.

Given two implementation of the micro-kernel, we vary the instruction tile
sizes and compare the performance. Our reference implementation  uses
uniformly sized N-updates of size $m_s \times n_s = 4 \times 2$. We compare this to
an implementation  composed of two types of N-updates of size $m_s \times n_s= 4 \times 3$
and $m_s \times n_s = 4 \times 1$. We illustrate these two implementations in
Figure~\ref{fig:nupdate_block_diagram}, where each outer-product is
partitioned according to a uniform or non-uniform scheme.

\begin{figure*}
  \includegraphics[clip, trim=1cm 2cm 1cm 2cm,scale=.5]{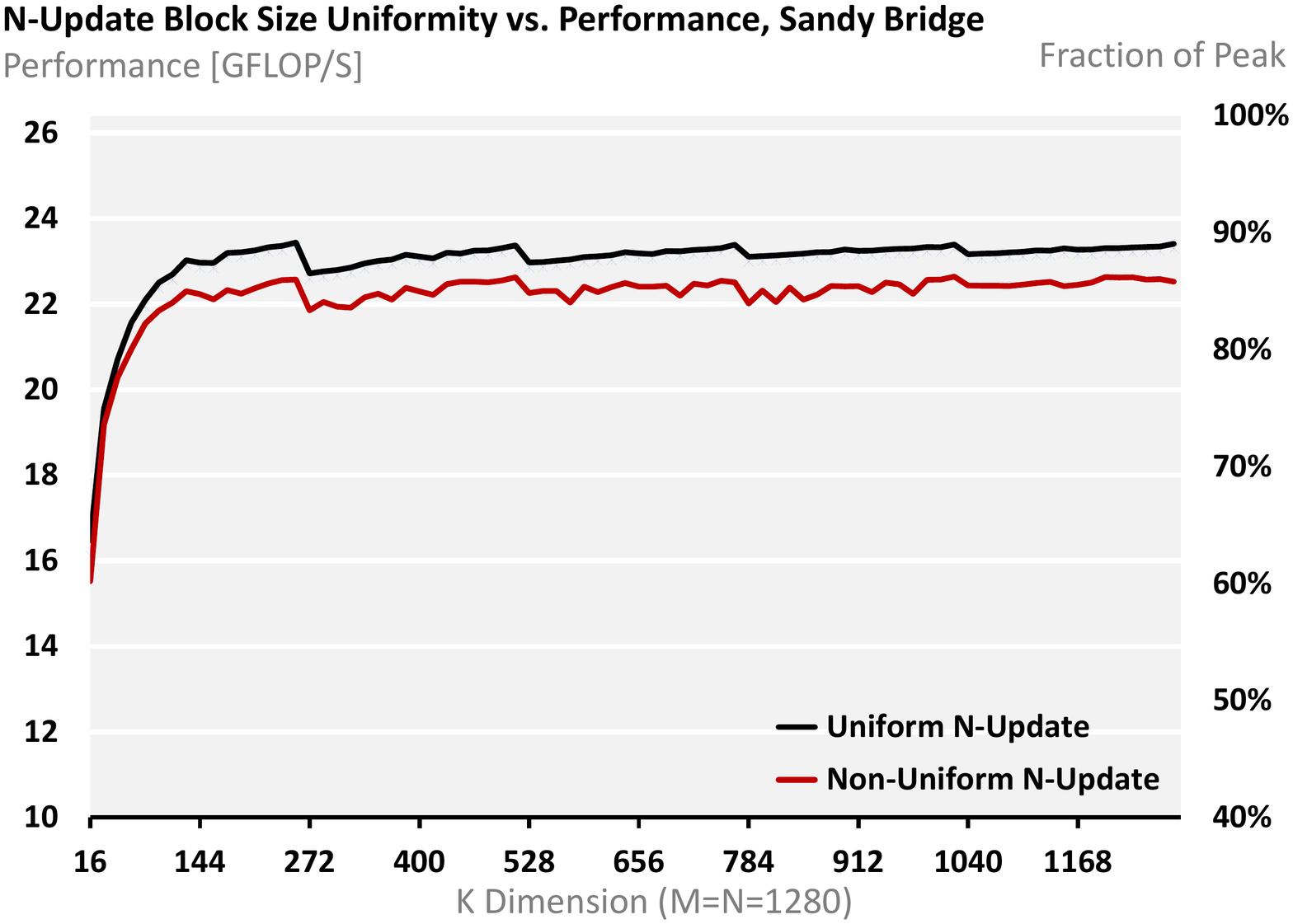}
  \caption{ In this experiment, we compare the performance of two kernels
    implementing the same  $m_r \times n_r = 8\times 4 $ outer-product using
    either uniform or non-uniform N-update sizes. The uniform N-update
    implementation performs better because it leads to few clusters of large
    (in Bytes) instructions which prevent the fetch and decode stages from
    becoming  bottlenecks.}
    \label{fig:uneven_experiment}
\end{figure*}

We ensure that both implementations are free of register
spilling and are scheduled --  not only to avoid stalls -- but also to insure
that the number of instructions between prefetch instructions and their
subsequent loads are uniform. We ran both implementations through the Intel
Code Architecture Analyzer (IACA), a software simulator for Intel
microarchitectures, and determined that both implementations lack instruction
stalls, spend an equal number of cycles on each functional unit, and have an
identical throughput (Figure~\ref{iaca_n_update}). However, the results in Fig~\ref{fig:uneven_experiment}
do not reflect the results we obtained from IACA because the non-uniform
N-update implementation performs 4\% worse than the uniform N-update.

The non-uniform N-update implementation leads to clusters of instructions with
very long encodings. This would present a bottleneck for the decoder and slow
down the overall execution rate. Using uniform N-updates results in large
instructions being evenly distributed throughout the code which prevents the
decoder from becoming a bottleneck.

\paragraph{Register}

For generating our kernels, our aversion to register spilling goes beyond the
performance penalty of the additional store to and load from memory. The reason
is that these kernels fit in a much larger {\Gemm } algorithm that achieves
high performance by reducing Translation Look-aside Buffer (TLB) misses, and by
spilling registers to memory this is disrupted and performance degrades
significantly as result of these TLB misses.

To demonstrate how large of an effect that register spilling in the kernel has
on the number of TLB misses, we evaluate three kernels with varying degrees of
register spilling (No Spilling, Moderate and Heavy). This is achieved by
varying how much the N-updates overlap when we schedule them using
software pipelining. The greater the overlap, the greater the register
pressure and the larger the number of spills. In addition to measuring
performance  (FLOPs per cycle), we also measure TLB misses using PAPI
\cite{Mucci99papi:a}. The goal is to show that by increasing the amount of
register spilling we will disrupt how the larger {\Gemm } algorithm avoids TLB
misses.

\begin{figure*}
  \includegraphics[clip, trim=1cm 2cm 1cm 2cm,scale=.5]{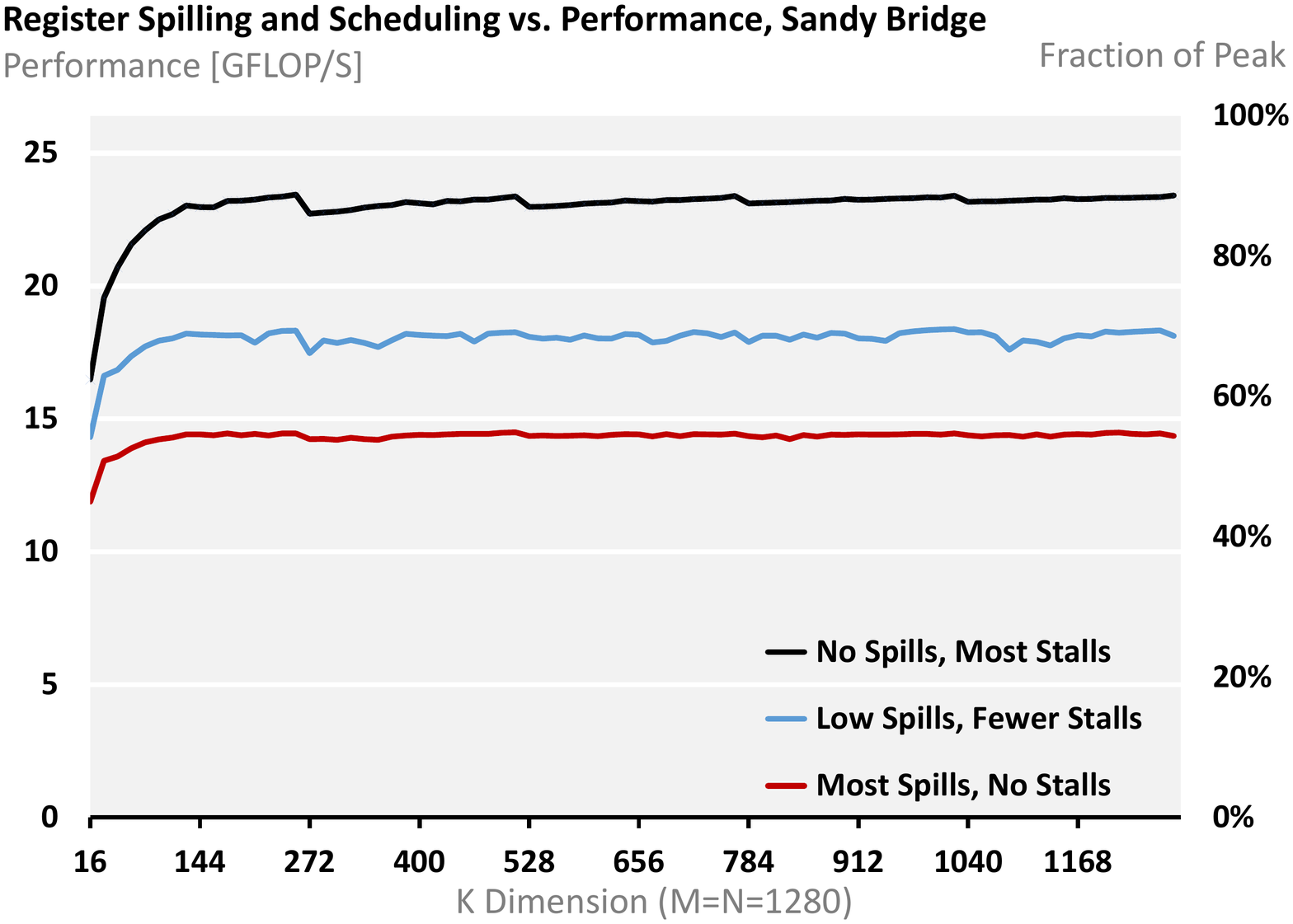}
  \label{fig:spilling_results}
  \caption{ In this experiment we show that improving static scheduling by
    increasing the number of register spills degrades the overall performance
    of the \Gemm operation. The GotoBLAS/BLIS algorithm attempts to minimize
    the number of TLB misses, however spilling into memory that would not have
    otherwise been used, increases TLB misses in this algorithm. }
\end{figure*}

The performance per cycle results in Fig~\ref{fig:spilling_results} demonstrate
that as we increase the number of spills performance decreases -- which is
what we would expect. We see that for large problem sizes the number of TLB
misses is greater for the Heavy amount of spilling compared to the Moderate
amount which is greater for the No Spilling case. If it were the case that the
added latency incurred by the register spills were the only source of
performance penalty, then we would not expect to see a change in the number of
TLB misses between the three cases.

This shows that register spilling has performance implication beyond the
additional round trip to cache because it disrupts the TLB miss avoiding
characteristics of the {\Gemm } algorithm described in \cite{Goto:2008:AHP}. For
practical purposes this removes spilling as an option when the outer-product
instruction-mix is translated into a kernel.

\begin{figure*}
  \includegraphics[clip, trim=1cm 2cm 1cm 2cm,scale=.5]{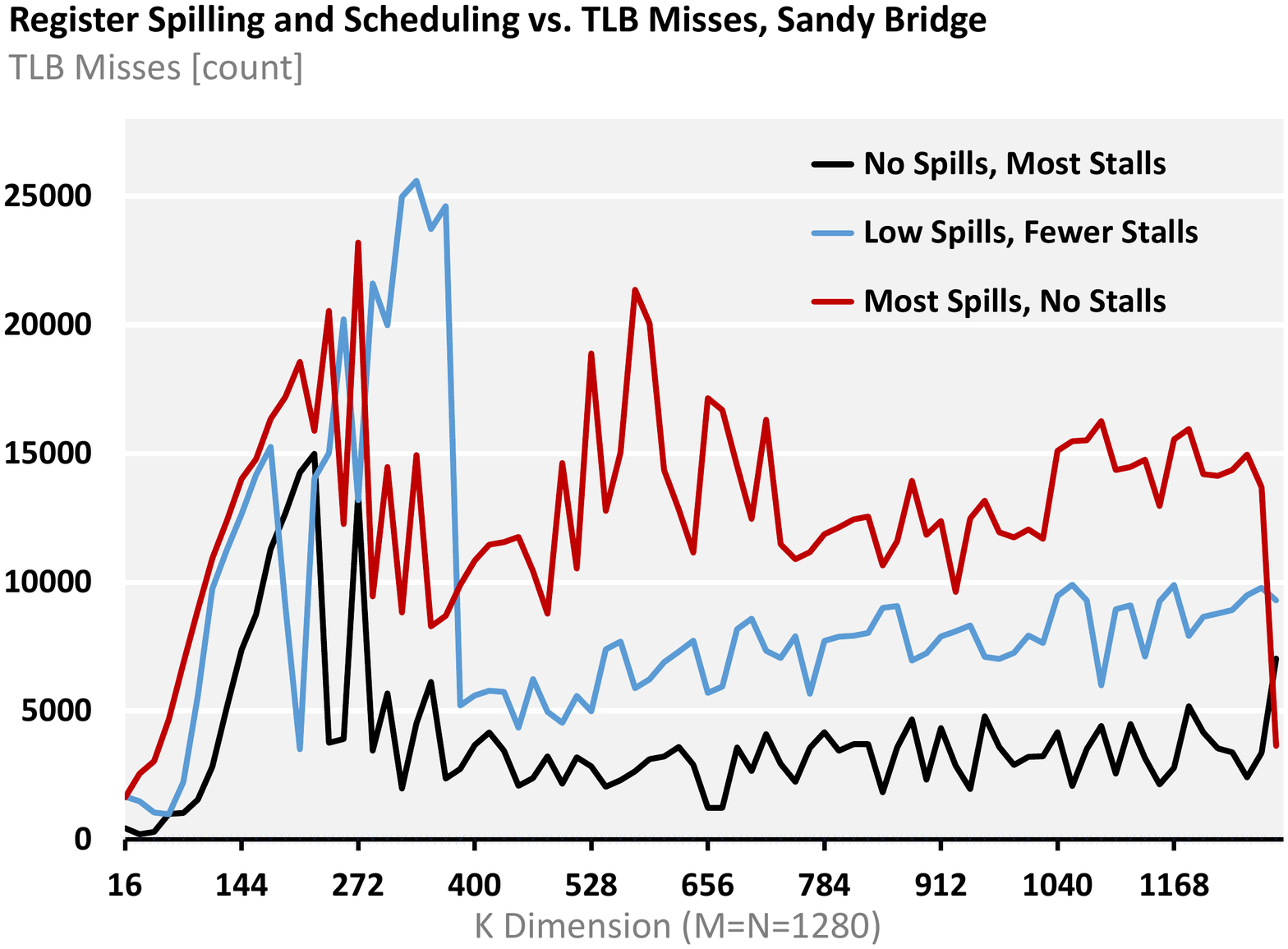}
  \caption{ The overall algorithm that our kernels embedded in, the
    GotoBLAS/BLIS \Gemm, maximizes performance by insuring that the kernel
    receives data at a sufficient rate while minimizing TLB misses. Spilling
    into memory requires that extra TLB entries be utilized to address memory
    that would not have been used otherwise. Thus even if spilling improves
    the kernel performance in isolation, it will degrade the overall
    performance of the {\Gemm } operation.}
\end{figure*}

\section{Conclusion}
In this paper, we address the last-mile problem of generating the
architecture-specific micro-kernel for the general matrix-matrix
multiplication routine that underlies most high performance linear
algebra libraries. Specifically, we reveal the system behind
generating high performance kernels that traditionally is implemented
manually by a domain expert.

We decompose the smallest unit of computation that is currently
written by the expert into even smaller units, which we term the unit
update. Using these unit update, we generate a family of algorithms
for computing the micro-kernel, and used analytical models to estimate
the running time for each possible implementation of the
micro-kernel. Finally, we generate one out of the many algorithms. We
demonstrate that our approach to systematically for generating this
micro-kernel yields kernels that are competitive with expert-optimized
versions.

To validate that the performance of our generated micro-kernels are
indeed high performance, we compared our generated results with
those from OpenBLAS, which uses a similar approach to high
performance matrix multiply. On many of the architectures, we
demonstrated that the generated kernels are within 2-5\% of the
OpenBLAS performance. In addition, we also show that the generated
kernels also scale in a similar fashion when parallelized on SMP
system such as the Xeon Phi. While analytically generating the micro-kernels makes us
competitive with expert-implemented kernels, empirical fine-tuning
could then be employed to recover the missing performance without
having to exhaustively search over a large space. This is something we
will explore in the future.


\bibliographystyle{acm}  
\bibliography{biblio}

\begin{thebibliography}{10}

\bibitem{LAPACK}
{\sc Anderson, E., Bai, Z., Bischof, C., Demmel, J., Dongarra, J., Du~Croz, J.,
  Greenbaum, A., Hammarling, S., McKenney, A., and Sorensen, D.}
\newblock {LAPACK:} {A} portable linear algebra library for high-performance
  computers.
\newblock In {\em Proceedings Supercomputing '90}. IEEE Computer Society Press,
  Los Alamitos, California, 1990, pp.~2--11.

\bibitem{PHiPAC97}
{\sc Bilmes, J., Asanovi\'c, K., whye Chin, C., and Demmel, J.}
\newblock Optimizing matrix multiply using \mbox{PHiPAC}: a {P}ortable,
  {H}igh-{P}erformance, \mbox{ANSI} {C} coding methodology.
\newblock In {\em Proceedings of International Conference on Supercomputing\/}
  (Vienna, Austria, July 1997).

\bibitem{BLAS3}
{\sc Dongarra, J.~J., Du~Croz, J., Hammarling, S., and Duff, I.}
\newblock A set of level 3 basic linear algebra subprograms.
\newblock {\em ACM Trans. Math. Soft. 16}, 1 (March 1990), 1--17.

\bibitem{Goto:2008:AHP}
{\sc Goto, K., and {v}an~de Geijn, R.}
\newblock Anatomy of high-performance matrix multiplication.
\newblock {\em ACM Trans. Math. Soft. 34}, 3 (May 2008), 12:1--12:25.

\bibitem{Goto:2008:HIL3}
{\sc Goto, K., and {v}an~de Geijn, R.}
\newblock High-performance implementation of the level-3 {BLAS}.
\newblock {\em ACM Trans. Math. Soft. 35}, 1 (July 2008), 1--14.

\bibitem{GMH:92}
{\sc Henry, G.}
\newblock {BLAS} based on block data structures.
\newblock Theory Center Technical Report CTC92TR89, Cornell University, Feb.
  1992.

\bibitem{inteloptimize}
{\sc {Intel Corporation}}.
\newblock {\em {Intel\textsuperscript{\textregistered} 64 and IA-32
  Architectures Optimization Reference Manual}}.
\newblock Sept. 2015.

\bibitem{poorman_journal}
{\sc K{\aa}gstr\"{o}m, B., Ling, P., and Loan, C.~V.}
\newblock Gemm-based level 3 blas: High-performance model implementations and
  performance evaluation benchmark.
\newblock {\em ACM Trans. Math. Softw. 24}, 3 (September 1998), 268--302.

\bibitem{LGEN2}
{\sc Kyrtatas, N., Spampinato, D.~G., and P\"{u}schel, M.}
\newblock A basic linear algebra compiler for embedded processors.
\newblock In {\em Proceedings of the 2015 Design, Automation \& Test in Europe
  Conference \& Exhibition\/} (San Jose, CA, USA, 2015), DATE '15, EDA
  Consortium, pp.~1054--1059.

\bibitem{software_pipeline}
{\sc Lam, M.}
\newblock Software pipelining: an effective scheduling technique for vliw
  machines.
\newblock In {\em Proceedings of the ACM SIGPLAN 1988 conference on Programming
  Language design and Implementation\/} (New York, NY, USA, 1988), PLDI '88,
  ACM, pp.~318--328.

\bibitem{little_law}
{\sc Little, J. D.~C.}
\newblock A proof for the queuing formula: L = $\lambda$ w.
\newblock {\em Operations Research 9}, 3 (1961), 383--387.

\bibitem{BLIS4}
{\sc Low, T.~M., Igual, F.~D., Smith, T.~M., and Quintana-Orti, E.~S.}
\newblock Analytical modeling is enough for high-performance blis.
\newblock {\em ACM Trans. Math. Softw. 43}, 2 (August 2016), 12:1--12:18.

\bibitem{Mucci99papi:a}
{\sc Mucci, P.~J., Browne, S., Deane, C., and Ho, G.}
\newblock {PAPI}: A portable interface to hardware performance counters.
\newblock In {\em In Proceedings of the Department of Defense HPCMP Users Group
  Conference\/} (1999), pp.~7--10.

\bibitem{OpenBLAS}
\url{http://xianyi.github.com/OpenBLAS/}, 2012.

\bibitem{Padua:1986:ACO:7902.7904}
{\sc Padua, D.~A., and Wolfe, M.~J.}
\newblock Advanced compiler optimizations for supercomputers.
\newblock {\em Commun. ACM 29}, 12 (Dec. 1986), 1184--1201.

\bibitem{BTOBLAS}
{\sc Siek, J.~G., Karlin, I., and Jessup, E.~R.}
\newblock Build to order linear algebra kernels.
\newblock In {\em 2008 IEEE International Symposium on Parallel and Distributed
  Processing\/} (April 2008), pp.~1--8.

\bibitem{BLIS3}
{\sc Smith, T.~M., {v}an~{d}e Geijn, R., Smelyanskiy, M., Hammond, J.~R., and
  Zee, F. G.~V.}
\newblock Anatomy of high-performance many-threaded matrix multiplication.
\newblock In {\em IPDPS '14: Proceedings of the International Parallel and
  Distributed Processing Symposium\/} (2014).
\newblock To appear.

\bibitem{LGEN}
{\sc Spampinato, D.~G., and P{\"u}schel, M.}
\newblock A basic linear algebra compiler.
\newblock In {\em International Symposium on Code Generation and Optimization
  (CGO)\/} (2014), pp.~23--32.

\bibitem{LGEN3}
{\sc Spampinato, D.~G., and P{\"u}schel, M.}
\newblock A basic linear algebra compiler for structured matrices.
\newblock In {\em International Symposium on Code Generation and Optimization
  (CGO)\/} (2016), pp.~117--127.

\bibitem{BLIS1}
{\sc Van~Zee, F.~G., and van~de Geijn, R.~A.}
\newblock Blis: A framework for rapidly instantiating blas functionality.
\newblock {\em ACM Trans. Math. Softw. 41}, 3 (June 2015), 14:1--14:33.

\bibitem{hands_off_hands_on}
{\sc Veras, R., Popovici, D.~T., Low, T.~M., and Franchetti, F.}
\newblock Compilers, hands-off my hands-on optimizations.
\newblock In {\em Proceedings of the 3rd Workshop on Programming Models for
  SIMD/Vector Processing\/} (New York, NY, USA, 2016), WPMVP '16, ACM,
  pp.~4:1--4:8.

\bibitem{AuGem}
{\sc Wang, Q., Zhang, X., Zhang, Y., and Yi, Q.}
\newblock {AUGEM}: Automatically generate high performance dense linear algebra
  kernels on x86 {CPUs}.
\newblock In {\em Proceedings of SC13: International Conference for High
  Performance Computing, Networking, Storage and Analysis\/} (New York, NY,
  USA, 2013), SC '13, ACM, pp.~25:1--25:12.

\bibitem{ATLAS}
{\sc Whaley, R.~C., and Dongarra, J.~J.}
\newblock Automatically tuned linear algebra software.
\newblock In {\em Proceedings of SC'98\/} (1998).

\bibitem{BLIS2}
{\sc Zee, F. G.~V., Smith, T.~M., Marker, B., Low, T.~M., Geijn, R. A. V.~D.,
  Igual, F.~D., Smelyanskiy, M., Zhang, X., Kistler, M., Austel, V., Gunnels,
  J.~A., and Killough, L.}
\newblock The blis framework: Experiments in portability.
\newblock {\em ACM Trans. Math. Softw. 42}, 2 (June 2016), 12:1--12:19.

\end{thebibliography}

\end{document}